\documentclass{aa}
\usepackage{graphicx}
\usepackage{txfonts}
\usepackage{natbib}
\bibpunct{(}{)}{;}{a}{}{,}    
\newcommand{\beq}{\begin{equation}}
\newcommand{\eeq}{\end{equation}}
\newcommand{\bea}{\begin{eqnarray}}
\newcommand{\eea}{\end{eqnarray}}

\newcommand{\subscr}[1]{_\mathrm{#1}}

\newcommand{\url}[1]{{\tt #1}}

\def\gapp{\lower 3pt\hbox{${\buildrel > \over \sim}$}\ }
\def\lapp{\lower 3pt\hbox{${\buildrel < \over \sim}$}\ }

\begin{document}
\title{On the evolution of eccentric and inclined protoplanets
embedded in protoplanetary disks}
\author{
Paul Cresswell \inst{2},
Gerben Dirksen \inst{1},
Willy Kley \inst{1}
\and
Richard P. Nelson \inst{2}
}
\offprints{W. Kley,\\ \email{wilhelm.kley@uni-tuebingen.de}}
\institute{
     Institut f\"ur Astronomie \& Astrophysik, 
     Universit\"at T\"ubingen,
     Auf der Morgenstelle 10, D-72076 T\"ubingen, Germany
\and
     Astronomy Unit
     School of Mathematical Sciences,
     Queen Mary, University of London,
     Mile End Road,
     London E1 4NS, United Kingdom
}
\abstract
{Young planets embedded in their protoplanetary disk
interact gravitationally with it leading to energy and angular momentum
exchange. This interaction determines the 
evolution of the planet through changes to the orbital parameters.}
{We investigate changes in the orbital elements of a
20 Earth--mass planet due to the torques from the
disk. We focus on the non-linear evolution
of initially non-vanishing eccentricity, $e$, and/or inclination, $i$.}
{We treat the disk as a two- or three-dimensional viscous
fluid and perform hydrodynamical simulations using finite difference methods.
The planetary orbit
is updated according to the gravitational torque exerted by the disk.
We monitor the time evolution of the orbital elements of the planet.}
{We find
rapid exponential decay of the planet orbital eccentricity and inclination 
for small initial values of $e$ and $i$, in agreement with linear theory.
For larger values of $e > 0.1$ the decay time increases
and the decay rate scales as $\dot{e} \propto e^{-2}$, consistent 
with existing theoretical models.
For large inclinations ($i > 6^\circ$) the inclination decay rate
shows an identical scaling $di/dt \propto i^{-2}$.
We find an interesting dependence of the migration on the eccentricity. 
In a disk with aspect ratio $H/r=0.05$ the
migration rate is enhanced for small non-zero eccentricities ($e < 0.1$),
while for larger values we see a significant reduction by a factor 
of $\sim 4$. 
We find no indication for a reversal of the migration for large $e$,
although the torque experienced by the planet becomes positive when
$e \simeq 0.3$. This inward migration is caused by the
persisting energy loss of the planet.
}
{For non gap forming planets,
eccentricity and inclination damping occurs on a time scale
that is very much shorter than the migration time scale.
The results of non linear hydrodynamic simulations are in very
good agreement with linear theory for values of $e$ and $i$
for which the theory is applicable
(i.e. $e$ and $i \le H/r$). }
\maketitle
\markboth
{Cresswell at al.: Eccentricity and Inclination of embedded planets}
{Cresswell et al.: Eccentricity and Inclination of embedded planets}
\section{Introduction}
\label{sec:introduction}
In the early stages of their formation protoplanets are still embedded
in the disk from which they formed. Not only will the protoplanet
accrete material from the disk, it will also interact gravitationally
with it.  Planet-disk interaction is an important aspect in
the process of planet
formation, and has been studied already before the discovery of
extrasolar planets through linear analysis
\citep{1980ApJ...241..425G, 1984ApJ...285..818P, 1986Icar...67..164W}.
In more recent times this was followed up by non-linear
numerical simulations using a two-dimensional setup
\citep{1999ApJ...514..344B, 1999MNRAS.303..696K, 1999ApJ...526.1001L,
2002A&A...385..647D}
and later in three dimensions as well
\citep{2003ApJ...586..540D, 2003MNRAS.341..213B, 2004A&A...418..325S}.

Since detection of the first extrasolar planet in 1995 there have
been about 200 additional discoveries (for an up-to-date list see
e.g. http://exoplanet.eu/ by J.Schneider). One of
the surprising differences between our own Solar System and these
extrasolar planets are their orbital properties.  In the Solar
System the larger (giant) planets move on almost circular orbits, 
whereas in contrast most extrasolar planets move on eccentric orbits. 
For a recent overview see
\citet{2005PThPS.158...24M}. 
The average orbital eccentricity of observed extrasolar planets is 
$e \simeq 0.3$. Not only are massive planets of several
Jupiter masses found in eccentric orbits,
but so are lighter planets in the mass range where they cannot open 
a clean gap in the disk ($M < M_{Jup}$).

One  possible origin of  the  eccentricity  is
through  interaction  with  the  protoplanetary disk.  In  particular,
\citet{2003ApJ...585.1024G}, \citet{2004ApJ...606L..77S}  estimate   that
eccentric Lindblad  resonances can  cause eccentricity  growth  for gap forming
planets. On the other hand, linear studies,
for example by 
\citet{1988Icar...73..330W}, \citet{2000MNRAS.315..823P} 
and \citet{2004ApJ...602..388T},
predict eccentricity damping for low mass embedded planets. 
Numerical hydrodynamical  simulations of embedded planets tend  to  show  
eccentricity damping as well.
For example \citet{2001A&A...366..263P} find  eccentricity growth for 
bodies on initially circular orbits, but only
for companions with a mass  greater than 10 $M_{Jup}$.
A recent study by \citet{2006ApJ...652.1698D}, however, suggests that modest
disk induced eccentricity growth can occur for planet masses
in the Jovian mass range. The origin of the on-average higher eccentricities
of the extrasolar planets remains to be definitively addressed, although
recent work indicates good agreement between planetary scattering
simulations and observational data \citep{2007astro.ph..3160J}.

The linear estimates of the eccentricity evolution of embedded planets
\citep{1993ApJ...419..166A, 1994Icar..110...95W, 2004ApJ...602..388T}
concentrate
on small eccentricities and predict exponential decay on short timescales
$\tau_{ecc} \approx (H/r)^2 \tau_{mig}$, 
where $H/r$ is the aspect ratio of the disk
and $\tau_{mig}$ and $\tau_{ecc}$ the migration 
and eccentricity damping time scale, respectively.
\citet{2000MNRAS.315..823P} have also considered larger values
for $e$ and they find an extended eccentricity 
damping time scale such that $de/dt \propto e^{-2}$
if $e > 1.1 H/r$. Additionally they find positive torques acting on the planet for
large values of $e$ and interpret this as outward migration.
Recently, \citet{2006A&A...450..833C} have performed hydrodynamical
simulations of embedded small mass planets and find good agreement with the work
by \citet{2000MNRAS.315..823P}.

The influence of the disk on the inclination of the planet has only been analysed in
linear studies by \citet{2004ApJ...602..388T}. They find exponential damping for
any non-vanishing inclinations on similar timescales as the eccentricity damping.
In both cases, their results are formally valid only for $e,i \ll H/r$, but numerical
experiments \citep{2006A&A...450..833C} suggest that at least their eccentricity 
damping estimates may be valid for a larger range of values and for a variety of
sub-gap opening planet masses.

Here we investigate the dynamical influence the protoplanetary disk has on a
protoplanet on an eccentric and/or inclined orbit.  
To model the disk dynamics we use a fully non-linear hydrodynamical description
in both two and three dimensions.
The orbit of the embedded planet is allowed to evolve due to the torques from the
disk. While in \citet{2006A&A...447..369K} the possibility
of eccentricity growth within the disk due to high mass planets 
($>$ 1 Jupiter mass)
has been studied, we now analyse the planet-disk interaction of a low-mass planet
of 20 Earth masses. 
Our work extends the recent two-dimensional analysis of \citet{2006A&A...450..833C}.
We vary the initial orbit parameters 
and analyse the influence of a nonzero eccentricity and inclination on the migration
rate of the planet. In particular we investigate if there are 
important differences
between the circular, the low eccentricity and the high eccentricity regimes.  
We analyse the time scale of the
eccentricity and inclination damping and compare to previous linear analysis.
We also compare the migration rates for the different orbital parameters with
the circular coplanar case and the analytical values from \citet{2002ApJ...565.1257T},
and determine an experimental upper bound on the validity of the exponential decay 
of eccentricity and inclination modelled by the analytical estimates of \citet{2004ApJ...602..388T}.

This paper is organised as follows. In section~\ref{sec:hydro-model}
we discuss the disk models and numerical methods.
In section~\ref{sec:eccentricity} we discuss the results for
planets on initially eccentric orbits located in the 
disk midplane. In section\ref{sec:inclination} we discuss the results
for planets on inclined and/or eccentric orbits.
Finally we discuss our results and draw conclusions 
in section~\ref{sec:conclusion}

\section{The Hydrodynamical Model}
\label{sec:hydro-model}
To study the evolution of an embedded planet in a protoplanetary disk we treat the
disk as a viscous fluid. 
We consider both 2- and 3-dimensional (2D, 3D) models which enables us to analyse the influence
of dimensionality directly.
Finding the right setup for the 2D models to agree with the 
more complex 3D results
will enable us to reduce computational efforts in the future.  
The origin of the coordinate system typically 
is star centred (in particular for the 2D simulations
and those 3D models presented in section~\ref{sec:inclination}),
but for some simulations it is located at
the centre of mass of the planet and star system.
The $z = 0$ plane defines the disk midplane and the planetary 
inclination is measured with
respect to this plane.
Many simulations were performed in a reference system that would be 
initially corotating with the planet if
it were on a circular orbit. This rotation rate is kept constant during the simulations
so that we do not have to consider extra accelerations.
Simulations presented in section~\ref{sec:inclination} were performed
in the inertial frame.

For the 2D models we use cylindrical coordinates ($r, \varphi, z$) and 
consider a vertically averaged, infinitesimally thin disk located
at $z=0$. 
The basic hydrodynamic equations (mass and momentum conservation)
describing the time evolution of such a
two-dimensional ($r, \varphi$) disk with embedded planets have been stated frequently and are
not repeated here \citep[see][]{1999MNRAS.303..696K}.
The 2D models presented here are calculated 
basically in the same manner as those described previously in
\citet{1998A&A...338L..37K, 1999MNRAS.303..696K} using the code {\tt RH2D}.
The reader is referred to those papers
for details on the computational aspects of this type of simulations.
Other similar models, following explicitly the motion of single
planets in disks, have been presented by
\citet{2000MNRAS.318...18N} and \citet{2000ApJ...540.1091B}.

For the 3-dimensional models we work in spherical coordinates ($r, \varphi, \theta$)
where in the third dimension we use a Gaussian density profile,
consistent with a locally isothermal equation of state.
We also set $H/r = const.$
In the vertical direction the grid extends over several scale heights.
This approach is the same as that described in \citet{2001ApJ...547..457K} 
and \citet{2003ApJ...586..540D}. 
\subsection{Initial Setup}
The 3D ($r, \varphi, \theta$) computational domain
consists of a complete ring of the protoplanetary disk centred on the
star.  The radial extent of the computational domain (ranging from
$r\subscr{min}$ to $r\subscr{max}$) is chosen such that there is sufficient
space on both sides of the planet to reduce possible influence of the
boundaries. Typically, we assume
$r\subscr{min}=0.4$ and $r\subscr{max}=2.5$ in units where the planet
is located initially at $r=1$.  In the azimuthal direction for a complete
annulus we have $\varphi\subscr{min} =0$ to $\varphi\subscr{max} = 2
\pi$. In the meridional direction we take a wedge spanning several disk scale
heights: $\theta_{min}=75^\circ$ and $\theta_{max}=105^\circ$. To save
computer time, the models without inclination were run in half a disk with
$\theta_{max}=90^\circ$, using symmetry around the midplane. 
Finally in the 2D simulations we use the same radial extent as in the 3D case.

The initial structure of the disk (density, temperature,
and velocity) is axisymmetric.
For the density $\rho(r, \varphi, \theta$) we assume a Gaussian stratification
in the vertical direction with a scale length $H$, and a power law for the 
radial dependence
\beq
      \rho(r, \varphi, \theta) = \rho_0 (r) \, \exp{[ - 1/2 (r\theta/H)^2]}
\eeq
where we choose $\rho_0(r) \propto r^{-1.5}$ such that the  
initial surface density $\sigma(r) = \int \rho dz$ falls off radially as
a power law with index -0.5. If there is no planet, this equilibrium
density structure will be preserved for constant kinematic viscosity and closed radial boundaries
\citep{2001ApJ...547..457K}.
The initial velocity is pure
Keplerian rotation ($u_r=0, u_\varphi = (G M_*/r)^{1/2}, u_\theta=0$).
The temperature stratification is always given by $T(r) \propto r^{-1}$
which follows from an assumed constant vertical height $H/r = const$.  For
these locally isothermal models the temperature profile remains fixed
and no energy equation is required.
We use a constant kinematic viscosity coefficient $\nu$.
 
To preserve a constant mass in the disk we use reflecting boundary
conditions at the inner and outer boundary. 
For testing purposes we apply also damping boundary conditions as described
in \citet{2006MNRAS.370..529D}. 
In the azimuthal direction, periodic boundary conditions
for all variables are imposed.
At the upper and lower meridional boundary we also use reflecting boundary
conditions to preserve the total mass. In those models where only the
upper part of the disk is considered we use symmetric boundary
conditions at the midplane to simulate a lower half of the disk that
is symmetric to the upper half.

\subsection{Model parameters}
\label{chap:setup}

The computational domain is covered by 131 $\times$ 388 $\times$ 40
($N_r \times N_\varphi \times N_\theta$) grid cells which are spaced
equidistant in radius, azimuth and polar angle.  
Half the number of cells are used in the meridional direction
when symmetry about the midplane is assumed.

The mass of the planet
relative to the mass of the star in the different models is 6$\times
10^{-5}$, which corresponds to a 20 $M_\oplus$ planet for a Solar mass
star. We allowed the planetary orbit to evolve as a result of the torque
exerted by the disk. In these models we assume a disk mass inside the
computational domain ($r =$ 2.08 to 13 AU for a Solar type star) 
of $7.0 \times 10^{-3}M_{\odot}$.

A constant dimensionless kinematic viscosity $\nu =10^{-5}$ is used for all models. 
This corresponds to $\alpha = 0.004$ for a planet at $a = 5.2$ AU,
a typical value for a protoplanetary disk. We set $H/r=0.05$.

We let the planet evolve to observe the orbital evolution of the
system. The motion of the planet is integrated using a fourth order
Runge-Kutta integrator where the time step size is given by the
hydrodynamical time step. 
As a test we have run pure 2 and 3-body problems of one star with one or two
planets under the same conditions as in the full
hydrodynamical evolution (i.e. identical timestep size)
and find that the relative total energy loss
is less than $6 \times 10^{-6}$ for $\approx 2,500,000$
time steps, which is equivalent to several thousand periods of the
planet. The forces from the disk
material are calculated in first order and are added when the planet
positions are updated. Disk self-gravity is not included.

In these models we did not take into account torques (forces) from disk
material that lies closer to the planet than $r_{torq} = 0.8-1.0\, R_{Hill}$
where the Hill-radius is given by 
\beq R\subscr{Hill} = a\subscr{p} \,
\left( \frac{q}{3} \right)^{1/3} \eeq
where $a_p$ is the actual semi-major axis of the planet orbit. 
For the calculations in section~\ref{subsec:dynamics} we use
a smoothed transition for the torque cutoff. 

\subsection{Numerical issues}
\label{sect:smooth}
We use two different codes, {\tt NIRVANA} and {\tt RH2D} for our
calculations.  The numerical method used for both utilises a staggered mesh,
spatially second order finite difference method based, where advection
is based on the monotonic transport algorithm
\citep{1977JCoPh..23..276V}.  Due to operator-splitting the code is
semi-second order in time. Details of the {\tt NIRVANA} code have
been described in \citet{1998CoPhC.109..111Z}, where both the London and
T\"{u}bingen groups have added their own improvements to the code.
Application of the two dimensional code {\tt RH2D}
to the embedded planet problem is described in \citet{1998A&A...338L..37K}.

The use of a rotating coordinate system in most of our runs
requires special treatment of
the Coriolis terms to ensure angular momentum conservation
\citep{1998A&A...338L..37K}.

In calculating the gravitational potential of the planet we use a
smoothed potential of the form 
\beq \Phi_P = -  \frac{G m_p}{(s^2 + \epsilon^2)^{1/2}} \eeq
where $s$ is the distance from the planet.
For the smoothing length $\epsilon$ of the potential we choose the diagonal length of one grid cell in 
three dimensions (this corresponds to 80 \% of the Hill sphere radius),
 and to simulate the three-dimensional environment as well as possible
in two dimensions we use $\epsilon = 0.6 H$.

The viscous terms, including all necessary tensor components,
are treated explicitly.

An important issue in complex numerical hydrodynamical simulations is the 
study of numerical convergence and consistency. While consistency of our 
difference
equations is satisfied through derivation of them from Taylor-expansions, the
question of convergence is typically addressed through resolution studies.
As described below, the main results of our simulations are obtained
by running two- and three-dimensional disk-planet
simulations over several hundreds of orbital periods. Since it is not possible
to perform resolution studies on all physical setups we present in the next
sections the results obtained with our standard resolution 
($131\times388\times40$ for
a 3D-setup with inclined planet). Additional resolution studies performed on a
representative sample are described in more detail in the Appendix, and
briefly at the relevant description of the results in the text.
\section{Eccentricity evolution}
\label{sec:eccentricity}

For the study of eccentricity damping of a planet 
embedded in a protoplanetary disk we
use here non-inclined orbits and distinguish two different regimes.
First we will investigate the regime of low non-zero
eccentricity and then we will look at high eccentricities similar to those observed
in exoplanetary systems.
The time variation $e(t)$ of the planetary eccentricity
is shown for different initial eccentricities $e_0$ in Fig.~\ref{fig:e3dx}. 

\subsection{Low initial eccentricity}

If we start the planet with a sufficiently low eccentricity ($e_0 \leq 0.1$) we
observe an {\it exponential decay} of the orbital eccentricity of the planet. 

\begin{figure}[ht]
\begin{center}
\rotatebox{0}{
\resizebox{0.98\linewidth}{!}{%
\includegraphics{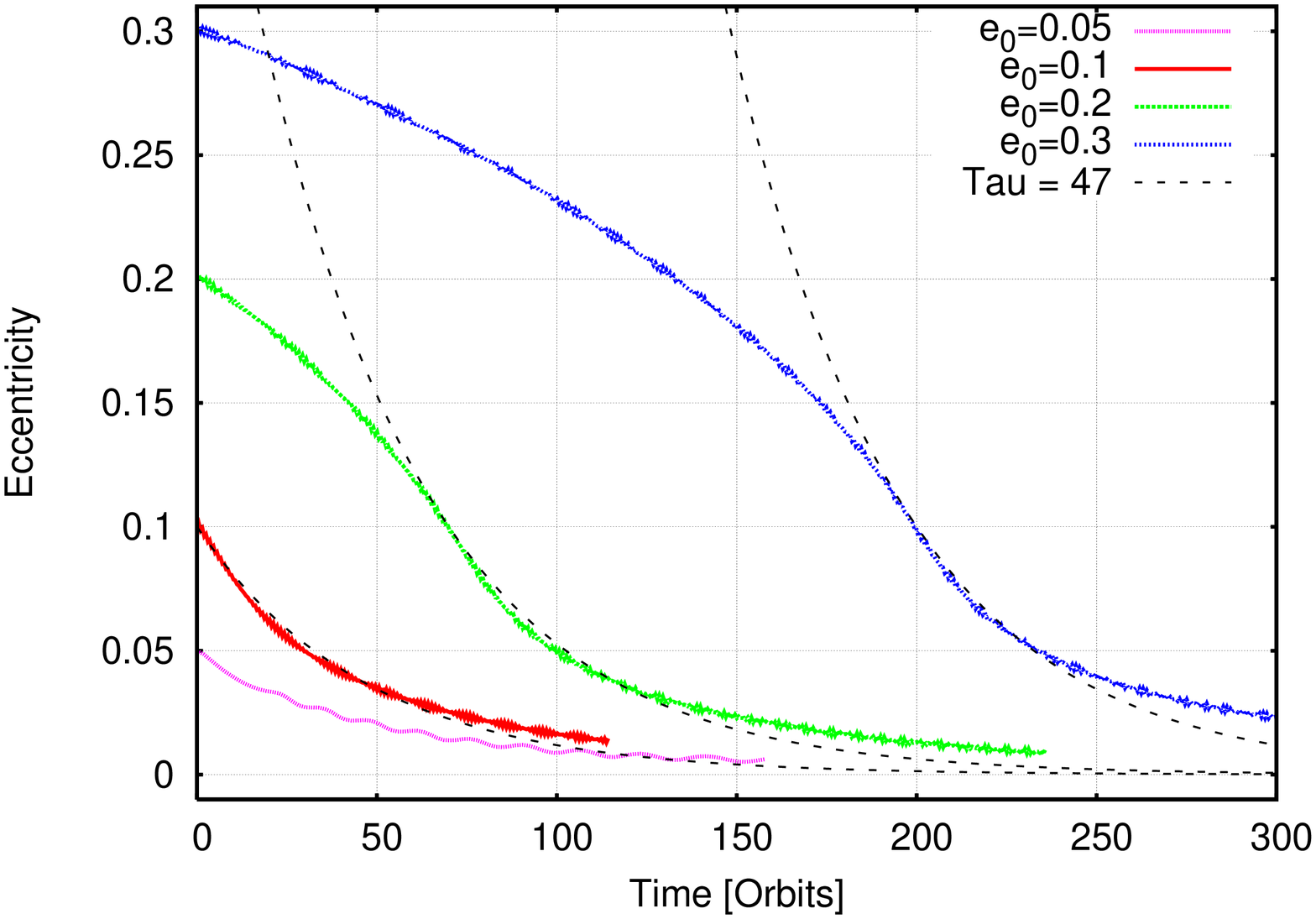}}}
\rotatebox{0}{
\resizebox{0.98\linewidth}{!}{%
\includegraphics{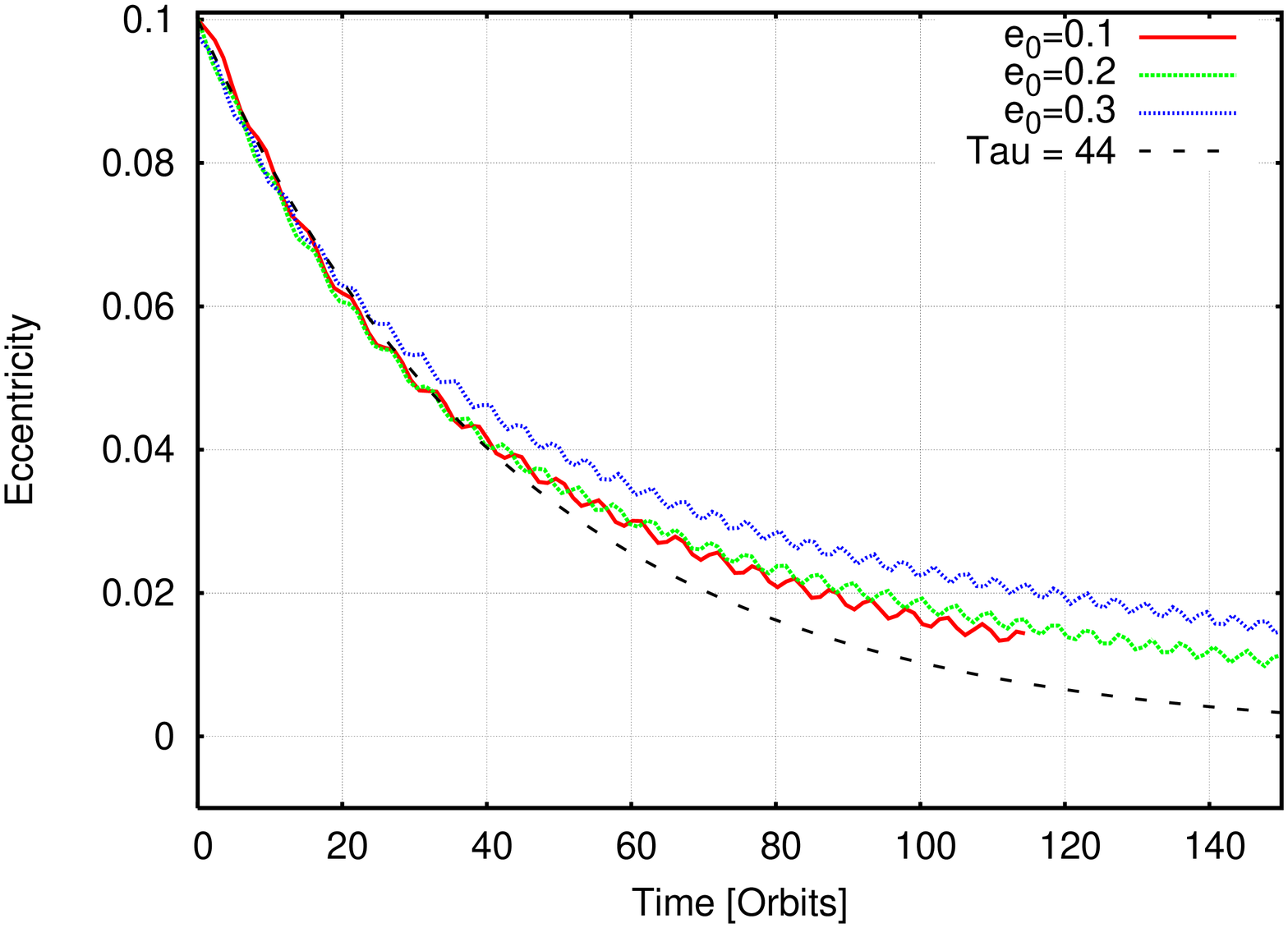}}}
\end{center}
  \caption{Eccentricity evolution of a 20 $m_E$ planet in a 3D protoplanetary disk as
a function of time for different initial eccentricities.
We find exponential decay in the low-eccentricity regime $e_0 \leq 0.10$, and slower damping
for higher eccentricities. {\bf Top:} The dashed lines indicate the exponential linear result
($\tau_{ecc}=47$) for this density as derived in \citet{2004ApJ...602..388T}.
{\bf Bottom:} Time-shifted results for the late exponential phase of the simulations
with large initial $e_0 \geq 0.1$. 
The dashed curve indicates here an exponential decay with $\tau_{ecc}=44$.
}
   \label{fig:e3dx}
\end{figure}

\begin{figure}[ht]
\begin{center}
\rotatebox{0}{
\resizebox{0.98\linewidth}{!}{%
\includegraphics{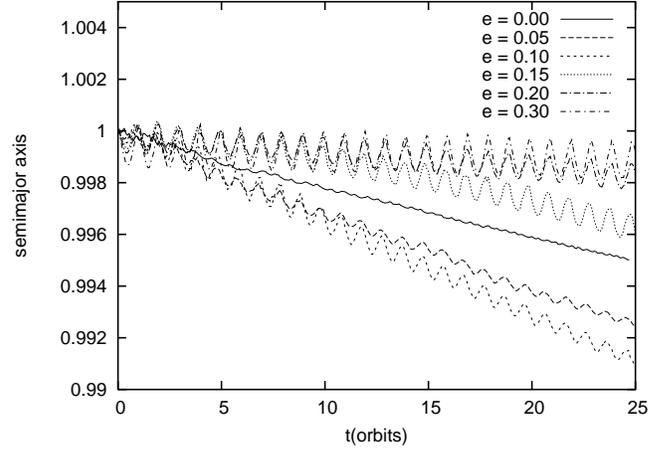}}}
\end{center}
  \caption{Semi-major axis evolution of the planet in a 3D protoplanetary disk
as a function of time for different initial eccentricities. 
For small non-zero eccentricities the migration rate increases above the $e=0$ case. 
For large eccentricities the migration rate is significantly slower.}
   \label{fig:alow}
\end{figure}

We find that for the low eccentricity models with $e_0 \le 0.10$ 
the decay time $t_{ecc} = |e/\dot{e}|$ is approximately 44 -- 50 orbits. 
Using linear analysis for small eccentricities 
\citet{2004ApJ...602..388T}
find that the mean eccentricity change
(averaged over one planetary orbit) is given by
\begin{equation}
\frac{\overline{de/dt}}{e} = - \frac{0.780}{t_{wave}}
\end{equation}
with characteristic time
\begin{equation}
\label{eq:twave}
t_{wave} = q^{-1} \left(\frac{\sigma_p a^2}{M_*}\right)^{-1}
\left(\frac{c_s}{a \Omega_p}\right)^4 \Omega_p^{-1}
\end{equation}
where $q$ is the mass ratio between the planet and the star, and
$\sigma_p$ the local surface density at the planetary orbit. For a
planet of 20 Earth masses at 5.2 AU in a MMSN nebula 
($\sigma_p = 1490\ {\rm kg~m^{-2}}$), 
as in our model, the characteristic time is $t_{wave}
= 438$ yrs $= 37$ orbits. This gives an eccentricity damping time scale
of about $\tau_{ecc} = t_{wave}/0.78 = 47$ orbits. This expected exponential
eccentricity evolution is indicated in Fig.~\ref{fig:e3dx} (upper panel)
by the dashed curves for $e_0 =0.1$ and (shifted) for $e_0=0.2$ and $0.3$
matched to the point where the curves cross $e=0.1$. 
In the lower panel of Fig.~\ref{fig:e3dx} we overlay the eccentricity
evolution for different $e_0$ such that they overlap at $e=0.1$.
The exponential decay phase is very similar for the different initial
eccentricities (see also below).
As seen from the plot our calculated $e$-damping time scale is only differs slightly,
being about 44 orbital periods for $e=0.1$ and increasing later as the eccentricity
damps further. This increase in damping time relative to that obtained
by \citet{2004ApJ...602..388T} is probably due to the influence of the gravitational softening
in our 3D simulations.

Overall, the linear estimates of \citet{2004ApJ...602..388T} are therefore 
in good agreement
with our numerical results for eccentricities $e \le 0.1$. Formally it
is expected that agreement should hold for $e \le H/r$, and for
our disks $H/r=0.05$, so the linear estimates appear to be good 
over twice their formal range of applicability.

As can be seen in Fig.~\ref{fig:alow} an eccentric orbit will also 
change the migration rate of the planet. The nearly straight solid line denotes the migration
for a planet on a circular orbit with $e_0=0.0$. 
Notice that in this case we always maintain the zero eccentricity.
For the eccentric planets the migration rate shows sinusoidal variations that
are caused by the orbital variations of the torques acting on the planet.

An interesting effect can be seen when one looks at the migration time
scale of the planet. For small initial eccentricities in the 
exponential damping regime,
here $e_0=0.05$ and $e_0=0.10$, we find a slightly faster 
migration rate for the planet than in the circular case.
This result is obtained for all codes used in this study,
and is found in both 2D and 3D runs. Its long term influence
on the migration rates of protoplanets is not significant, however,
as the eccentricity damping time scale is shorter than the migration time
by a factor on the order of $(H/r)^2$.

\begin{figure}[ht]
\begin{center}
\rotatebox{0}{
\resizebox{0.98\linewidth}{!}{%
\includegraphics{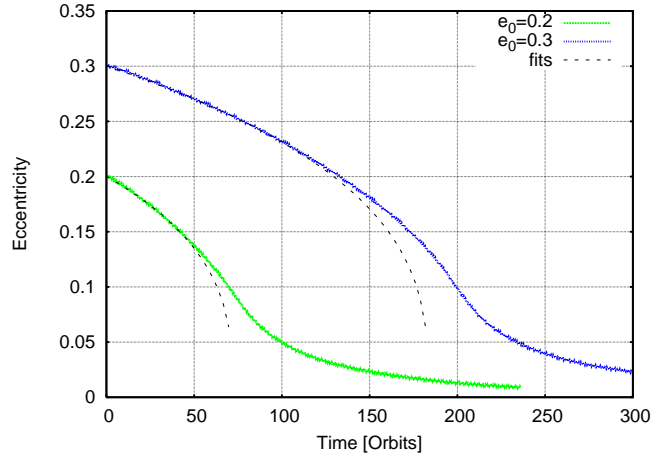}}}
\end{center}
  \caption{The models with $e_0 = 0.2$ and $0.3$ are fitted to a theoretical model
    with $\dot{e}\propto{e^{-2}}$. As can be seen this scaling
    works extremely well as long as the eccentricity is larger than
    about 2.5 $H/r$.  }
   \label{fig:efit}
\end{figure}

\subsection{High initial eccentricity}

If we start the planet with a high eccentricity ($e_0 > 0.1$) we also
observe a decay of the orbital eccentricity of the planet, but it is
slower and the decay rate is no longer exponential,
as can be seen in the two upper curves in Fig.~\ref{fig:e3dx}. 
In fact it fits well with the theoretical model 
described in \citet{2000MNRAS.315..823P},
which predicts $\dot{e}\propto e^{-2}$.  In Fig.~\ref{fig:efit} the
calculations with $e_0 = 0.2$ and $0.3$ are fitted with such a
model. The fits are extremely good. 
The characteristic damping time at $t=0$ for this non-exponential damping
with $\dot{e} = K e^{-2}$ is given by
\begin{equation}
    \tau_e = \left. \frac{e}{\dot{e}}\right|_{t=0} \, = \, \frac{e_0^3}{K}
\end{equation} 
From Fig.~\ref{fig:efit} we obtain $\tau_e = 71$ for $e_0 = 0.2$,
and for $e_0 = 0.3$ we find $\tau_e  \approx 183$ orbital periods, 
more than a four-fold increase over the exponential decay.
Also notice that when the eccentricity falls beneath the value
for which exponential damping occurs, the lines in  Fig.~\ref{fig:e3dx}
(lower panel) look similar to
shifted copies of each other, indicating that the planet
has no memory of its previously higher eccentricity.
In the large $e_0 =0.3$ case the additional evolution in semi-major
axis may account for the observed (slight) difference in decay time.

\begin{figure}[ht]
\begin{center}
\rotatebox{0}{
\resizebox{0.98\linewidth}{!}{%
\includegraphics{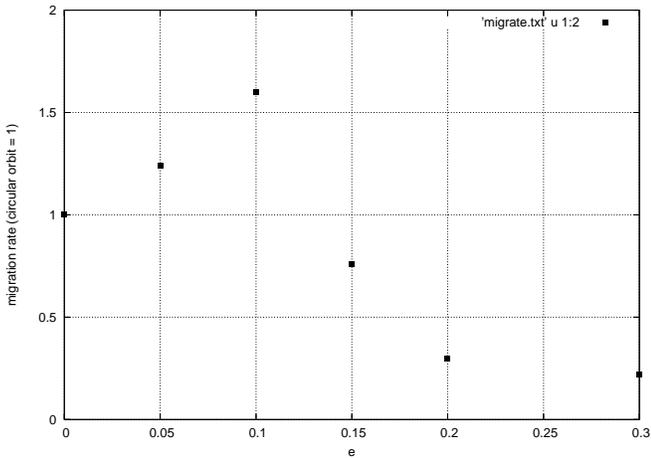}}}
\end{center}
  \caption{The relative initial migration rate as a function of eccentricity
  for a planet with mass ratio $q = 6 \times 10^{-5}$. 
  For low eccentricities the
  migration rate can increase by up to 60\%, for high eccentricities
  we find a substantial slowing of the migration.}
   \label{fig:migfunec}
\end{figure}

For these high initial eccentricities we find additionally a reduced migration rate
(Fig.~\ref{fig:alow}).
In Fig.~\ref{fig:migfunec} the relative migration rate compared to the
$e = 0$ model is plotted. For small eccentricities the migration rate
may {\it increase} by as much as 60\% before dropping off in
the manner expected for
larger eccentricities. For a large eccentricity of $e = 0.3$ the
migration rate is 5 times smaller than for a planet on a
circular orbit.

\begin{figure}[ht]
\begin{center}
\rotatebox{0}{
\resizebox{0.98\linewidth}{!}{%
\includegraphics{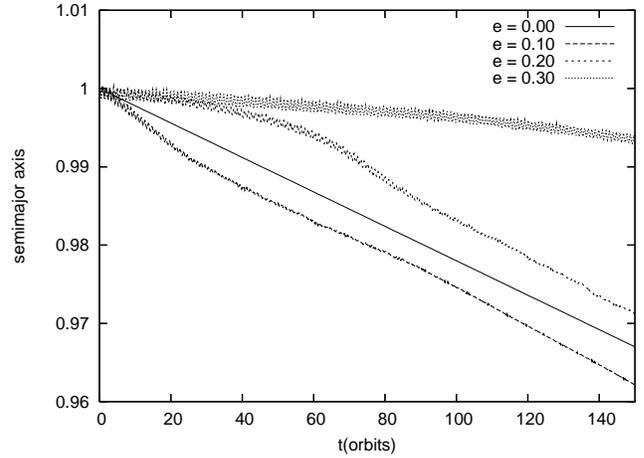}}}
\end{center}
  \caption{The long term evolution of the semi-major axis for different
  initial eccentricities for a 20 $m_E$ planet for 3D models. 
  All models except the one with the largest
  initial eccentricity converge to the same migration rate,
  and this large $e$ run eventually converges also after a longer
  run time.}
   \label{fig:semi}
\end{figure}

In the long term evolution of the migration rate we see in 
Fig.~\ref{fig:semi} 
that all models converge to the circular migration rate
when approaching $e = 0$, as expected. 
This turnover to the circular migration rate occurs at a residual eccentricity
of about $e=0.05 \approx H/r$, i.e. for the model with $e_0=0.10$ at around $t= 20$ and for
the model with $e_0=0.2$ at around $t=80$. 
For the $e = 0.3$ model the
eccentricity has not dropped enough to reach the turnover from the
slow migration regime (large $e$) to the standard migration regime (small
nonzero $e$), but will eventually reach the same migration rate as the
other models in Fig.~\ref{fig:e3dx} 
(see also the two-dimensional models in section \ref{sect:20e2d}).

The turnover from the slower migration rate for larger initial $e_0$
to the circular rate occurs through a brief phase of rapid migration
when the actual eccentricities are in a range of $0.1$ to $0.05$.
This occurs for the $e_0=0.20$ model during the time $t=60$ to $t=80$.

\subsection{Comparison with two-dimensional models}
\label{sect:20e2d}
\begin{figure}[ht]
\begin{center}
\rotatebox{0}{
\resizebox{0.98\linewidth}{!}{%
\includegraphics{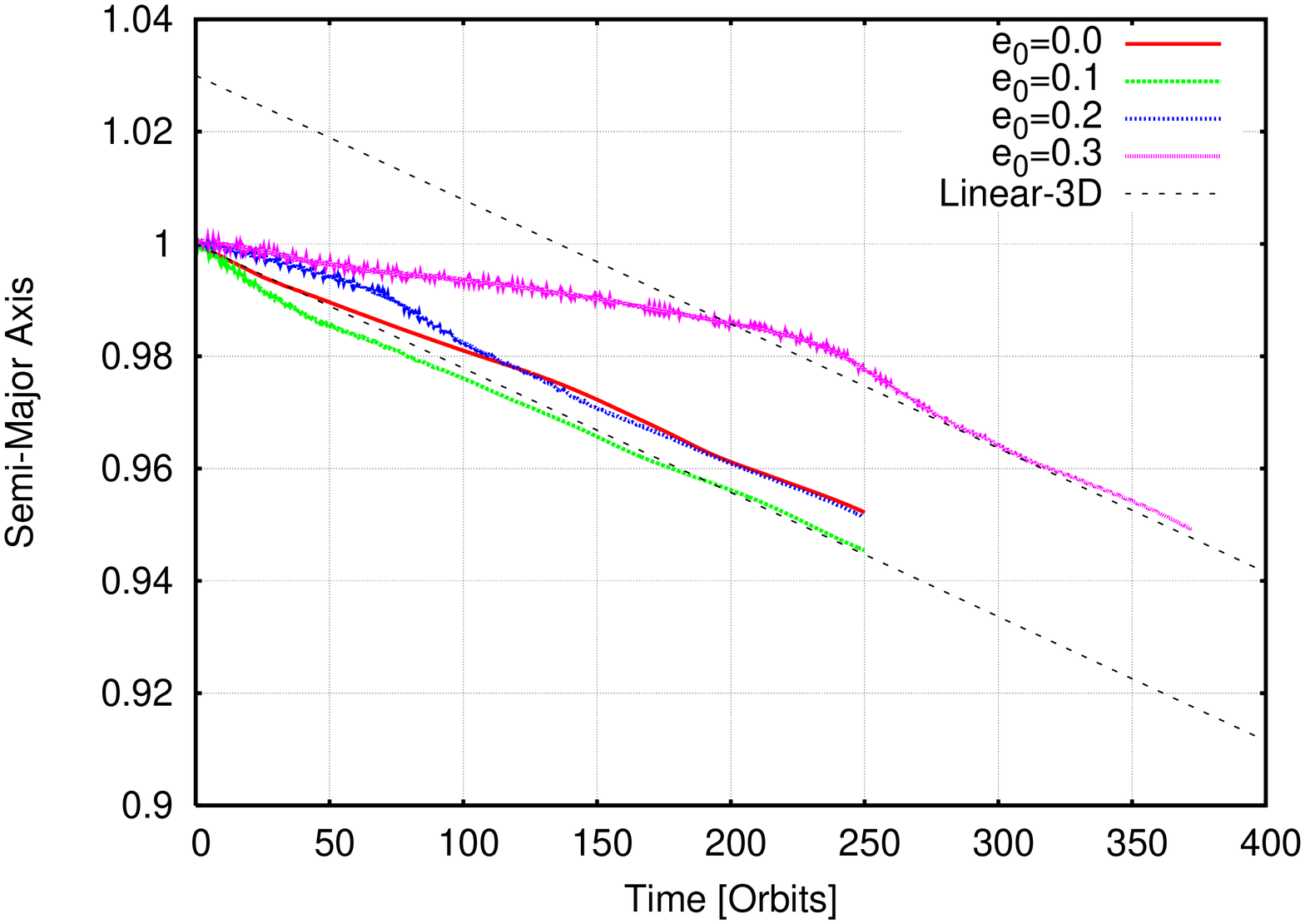}}}
\rotatebox{0}{
\resizebox{0.98\linewidth}{!}{%
\includegraphics{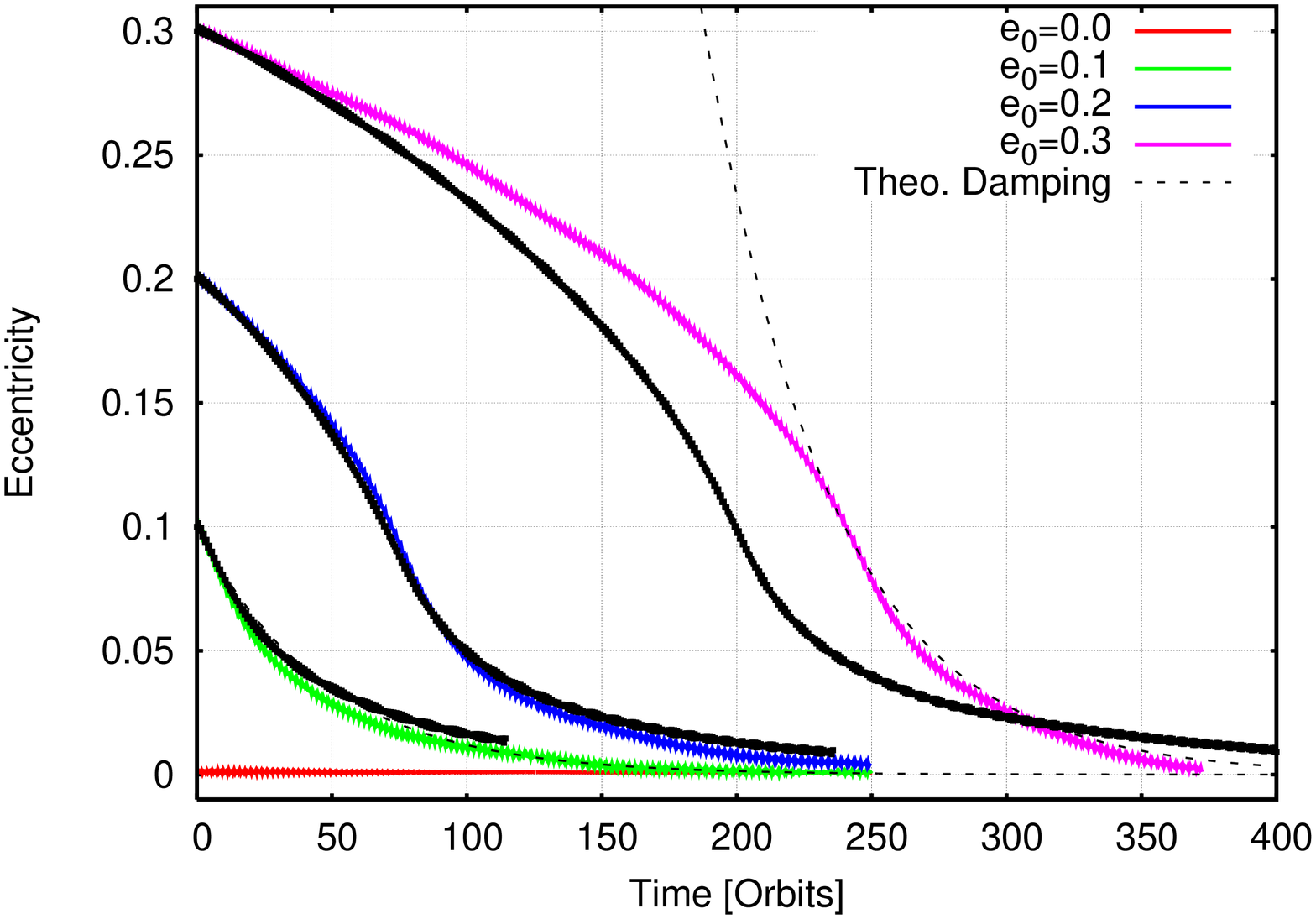}}}
\end{center}
  \caption{ {\bf Top:} Semi-major axis as a function of time for $q =
  6 \times 10^{-5}$ in a 2D disk for different initial eccentricities.
  For very high
  eccentricities the speed of migration is significantly reduced
  as in the 3D case. Dashed lines indicate the result from linear
   analysis for 3D disks. {\bf Bottom:}
  Eccentricity as a function of time for the same runs.
  The dashed lines indicate the linear results from \citet{2004ApJ...602..388T}
  with $\tau_{ecc}=47$. 
  Our 3D results are over-plotted using the small black symbols.
  For larger initial eccentricities 2D and 3D runs begin to differ; by about
  20\% for the $e_0=0.3$ case. Possible reasons for this behaviour
  are discussed in the text.
 }
\label{fig:2decc}
\end{figure}

To investigate the influence of dimensionality, we have
rerun all the previously described
models with a low mass eccentric planet in a
two-dimensional (2D) disk as well. Our initial conditions for the disk
are identical 
to the models in three dimensions, only vertically averaged. The grid
structure is the one described in section \ref{chap:setup}. The evolution
of the semi-major axis and eccentricity of a 20 Earth-mass planet 
starting on an
eccentric orbit with initial eccentricity between 0 and 0.3 for the
two-dimensional models is shown in Fig.~\ref{fig:2decc}. In the top
panel the semi-major axis is shown. 
When the potential smoothing described in section \ref{sect:smooth}
takes the value $\epsilon = 0.6 H$, the
migration rate in the two-dimensional disk agrees well with
for the migration rate in three dimensions obtained using linear
theory. The dashed reference lines refer to these linear
3D values as given by \citet{2002ApJ...565.1257T}.
The obtained migration rate agrees with our 3D-results as shown above.
Similarly to the 3D case, the models with small eccentricities $e \leq 0.10$ migrate
initially at a faster rate, while those for larger $e$ are migrating at a slower
rate.
Another
feature that is preserved from the three-dimensional calculations is
the sharp change in migration rate when the eccentricity drops below
$\approx$ 0.10. For the $e_0 = 0.30$ model this happens after approximately 240
orbits, for the $e_0 = 0.20$ model after approximately 70 orbits.
As described above, at this point during the evolution the eccentricity damping
changes from the slower $\dot{e} \propto  e^{-2}$ behaviour to the faster
exponential behaviour during which migration is slightly accelerated.
Only when the eccentricities become very small the migration is slowed again and
the standard (circular) rate is approached for all models. 

There are also some distinct differences. 
In the bottom panel of Fig.~\ref{fig:2decc}
the eccentricity is plotted as a function of time. 
Over-plotted (dashed curves) are the linear damping rates by 
\citet{2004ApJ...602..388T}, where our damping is again slightly faster.
Our damping rates for the 2D and 3D cases agree very well for the smaller
eccentricities while for the large value $e_0=0.3$ the 3D run gives a significantly
faster damping. 
However, the generic features and scaling of the damping process for
large $e_0$ is captured correctly in both 2D and 3D runs, and the
differences in the eccentricity damping are at the 20\% level even for
the largest $e_0$.

To examine the origin of this discrepancy, we performed additional 2D and 3D 
models with larger disks, $0.25 \le r \le 3.5$, while preserving grid
resolution, to rule out the possibility that boundary effects 
are the cause. 
A small difference in
eccentricity reduction was observed at the transition to linear damping, 
$de/dt \propto e$, with longer non-linear and shorter linear $e$-folding times
than for our standard disk model, but the difference is only a few percent 
and not of the order observed in Fig.~\ref{fig:2decc}.
Runs with enhanced damping at the radial boundaries to reduce reflections
did not change the results considerably, some damping is clearly required however,
to avoid unphysical influence due to the boundaries.

As a next step we analysed the role of potential softening in more detail.
While the use of a larger softening ($\epsilon = 0.6 H$) to approximate 3D disk
structure is a well-known approach in 2D modelling, for sufficiently eccentric
orbits in which a large region of the disk is sampled, 
our results indicate that a simple prescription for potential
softening may no longer be sufficient to obtain agreement 
between 2D and 3D simulations. 
This may be due in part to the fact that for highly eccentric orbits
the planet is crossing the resonances that drive the disk-planet interaction,
such that a strong sensitivity to the softening prescription
is expected \citep{2000MNRAS.315..823P}. Further calculations that
we have performed indicate that a considerably more complex
prescription for softening the gravitational potential
will be required if 2D models are to be constructed that agree with
3D models for both high and low orbital eccentricities.
Clearly, a softening that is spatially fixed and does not vary with radius
(which $H$ does in fact) will not be correct, a conclusion which applies
to the 3D case as well in case the simulation is resolution limited.
In our case at $r=1$ the standard grid resolution yields a diagonal 
length of $\Delta \approx 0.9 R_{Hill}$ for one gridcell.
The torque cutoff $r_{torq}$ should also scale with the radial
distance of the planet from the star and not be a function of semi-major
axis alone.  
\begin{figure*}[ht]
\begin{center}
\rotatebox{0}{
\resizebox{0.98\linewidth}{!}{%
\includegraphics{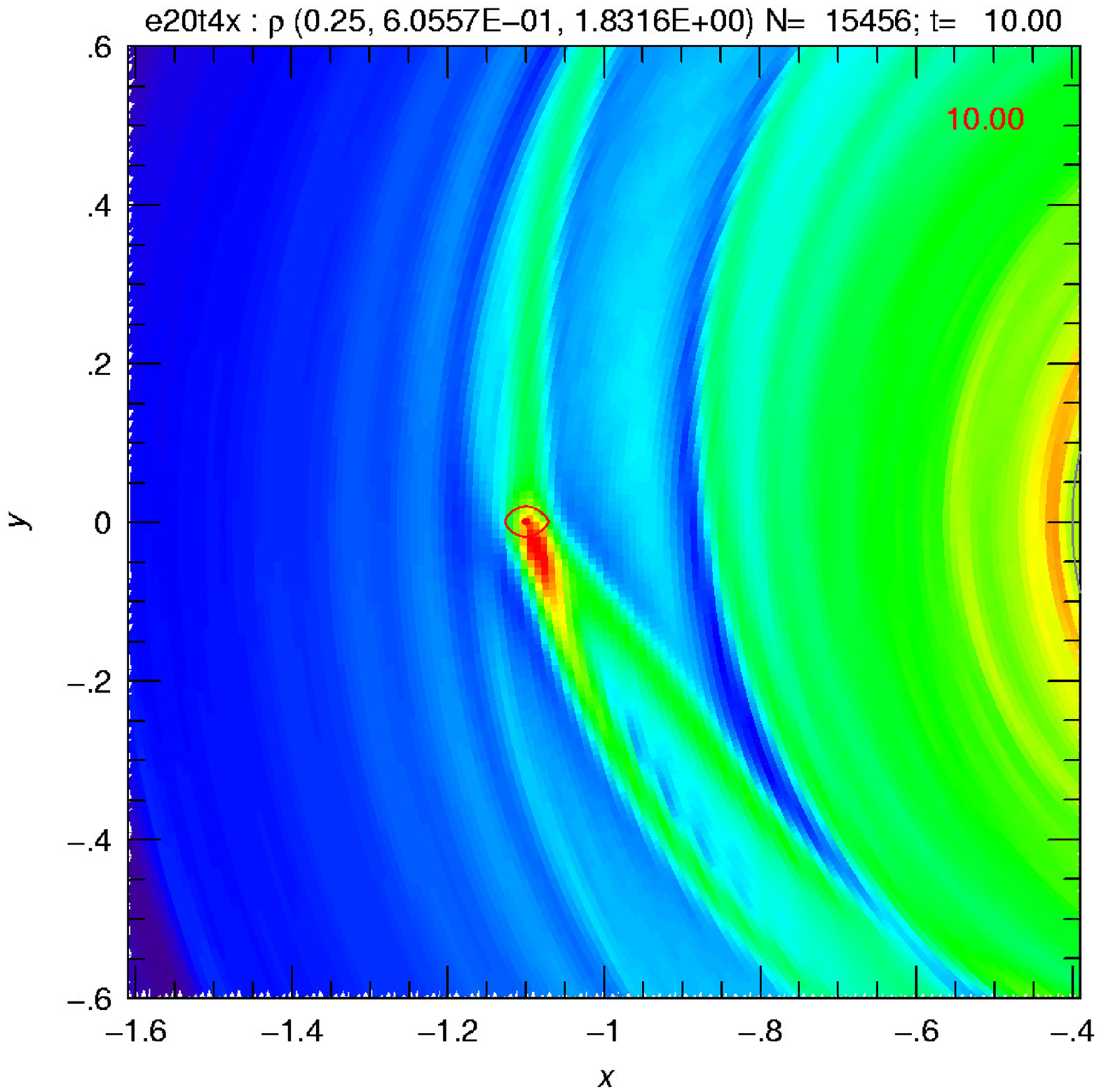}
\includegraphics{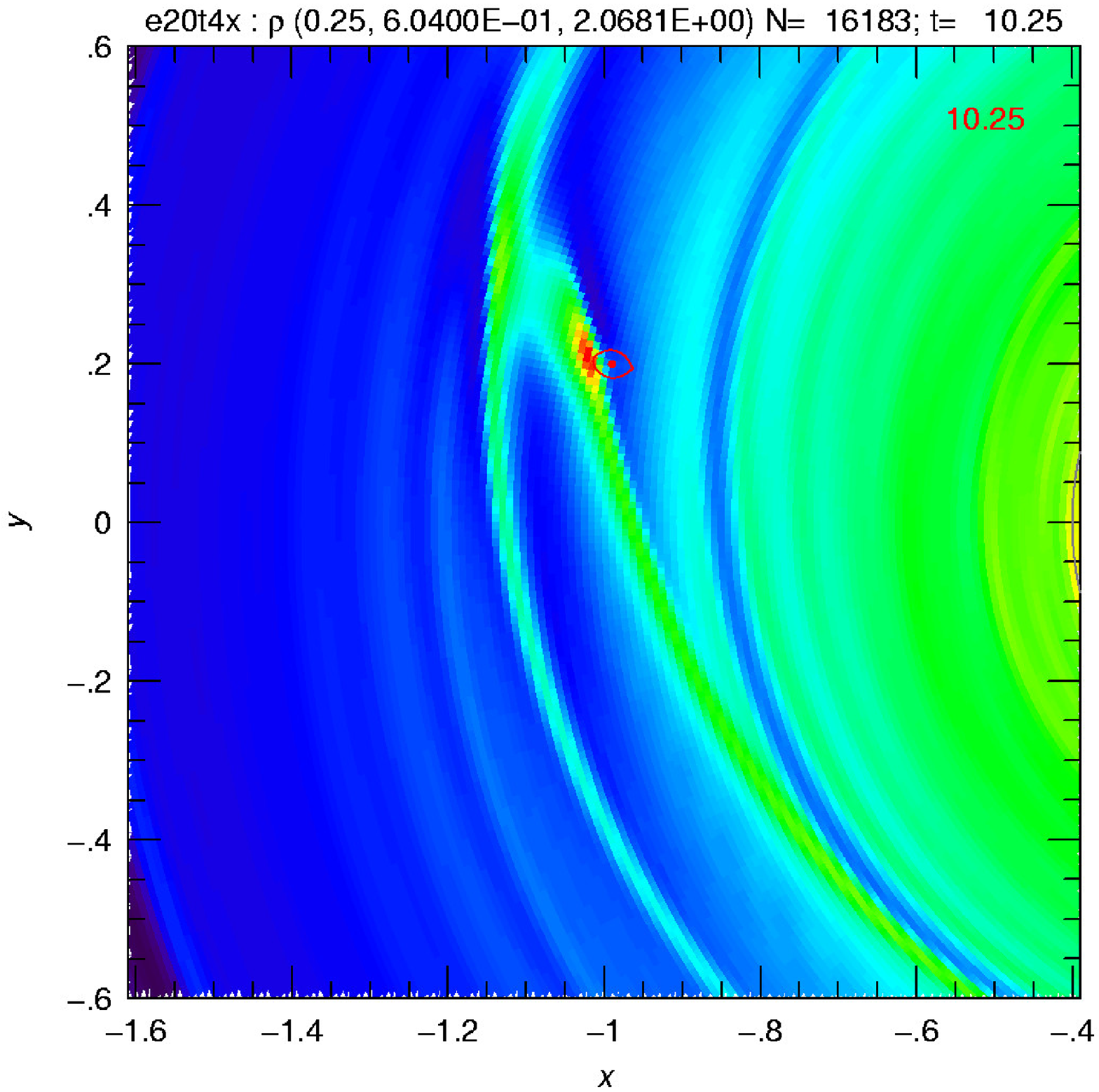} 
}} 
\rotatebox{0}{
\resizebox{0.98\linewidth}{!}{%
\includegraphics{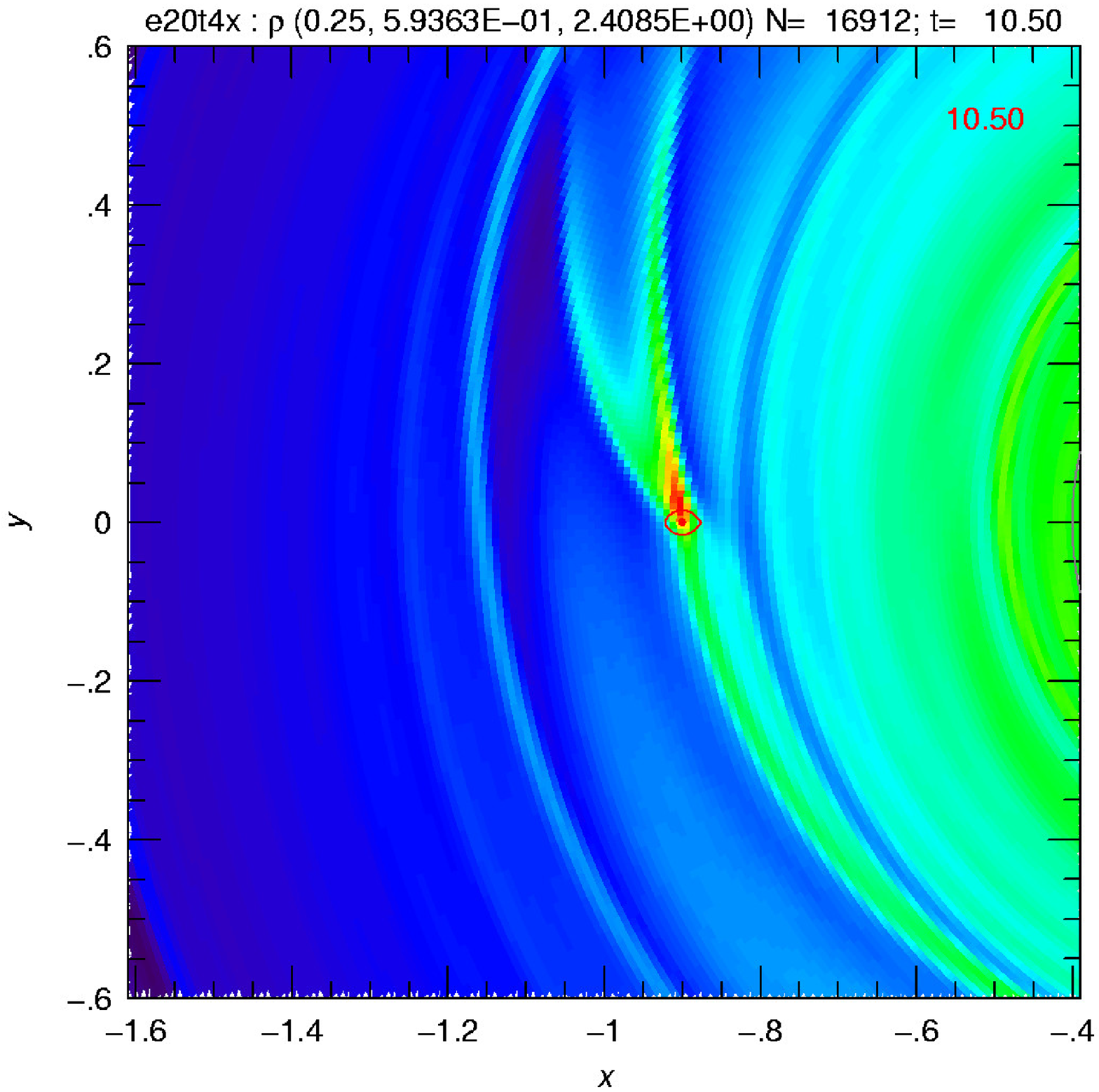}
\includegraphics{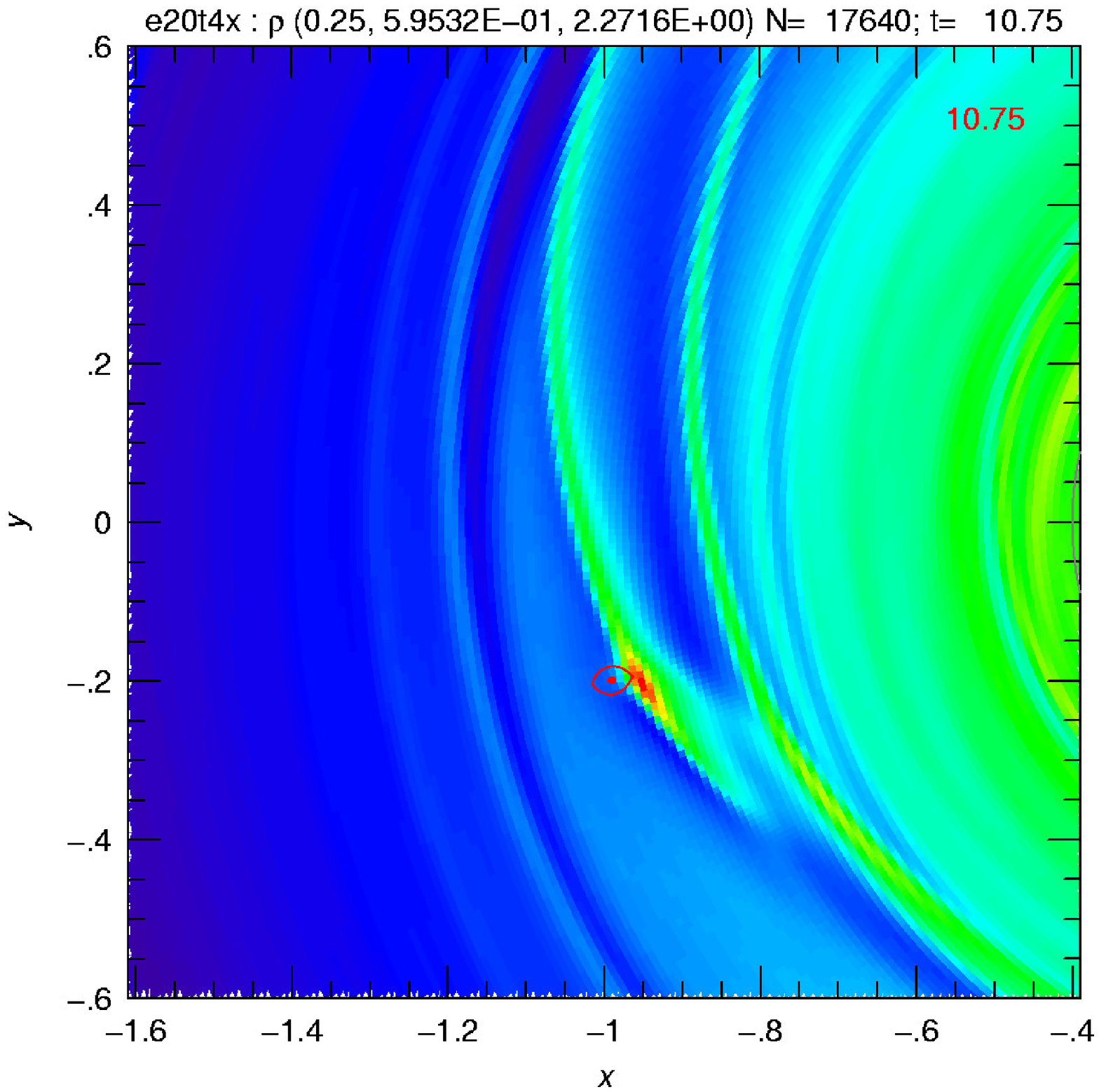}
}} 
\end{center}
  \caption{Density contour plot for a 20 Earth-mass planet on an fixed $e=0.1$ orbit
  imbedded in a 2D disk at four different times during one orbit.
  The snapshots are separated by 1/4 orbits.
 }
   \label{fig:2dflow}
\end{figure*}
We would like to point out that increasing the resolution of our models
in the two- and three-dimensional case changes the time evolution
of the semi-major axis only marginal, but lead to a slightly faster
eccentricity damping. However, the observed discrepancy in timescales between the 
2D and 3D case is not altered. See Appendix for a more detailed discussion
on resolution issues.
\subsection{Dynamics of the flow}
\label{subsec:dynamics}
To analyse the dynamical structure of the flow we consider 
the change of the flow field in the case of an eccentric planet for
a 2D case.
In Fig.~\ref{fig:2dflow} we display the density structure of the
disk with an embedded $q=6 \times 10^{-5}$ planet
on an eccentric orbit with fixed $e=0.1$ at four different phases in the orbit,
separated by 1/4 orbits, ranging from $t=10$ to $t=10.75$. 
At $t=10$ the planet is at apoastron (top left)
and at $t=10.5$ at periastron (bottom left).
In contrast to the evolution with zero eccentricity where two
stationary trailing arms exist, one in the outer disk ($r>r_p$) and one in the inner
disk ($r<r_p$), in this eccentric case spiral arms and additional flow features
appear and disappear periodically in phase with the orbit.
During periastron a pronounced  outer spiral attached to the planet
is visible while at apoastron this has changed to an inner spiral.
This periodic shift of the spiral arm strengths will result in a
corresponding periodic variation of the torque acting on the planet.
We note also that a significant density enhancement appears
in the close vicinity of the planet. At apoastron it clearly
lies in front of the planet (and thus exerts a strong positive torque),
and at periastron it lags behind the planet (exerting a strong
negative torque). This flow feature appears to arise because when
the orbit is eccentric the flow in the planet vicinity becomes similar
to a Bondi-Hoyle flow. At apoastron the planet moves more slowly
through the gas, and so is over-taken by the disk matter on
trajectories that lie both inside and outside the planet orbit.
These flow lines are distorted by the gravitational
field of the planet and come to a focus in front of the
planet, forming the high density feature seen in the top left panel
of Fig.~\ref{fig:2dflow}. The same effect happens in reverse
at periastron, leading to a high density feature that forms
behind the planet. These features become dominant in
determining the torques experienced by the planet.

\begin{figure}[ht]
\begin{center}
\rotatebox{0}{
\resizebox{0.98\linewidth}{!}{%
\includegraphics{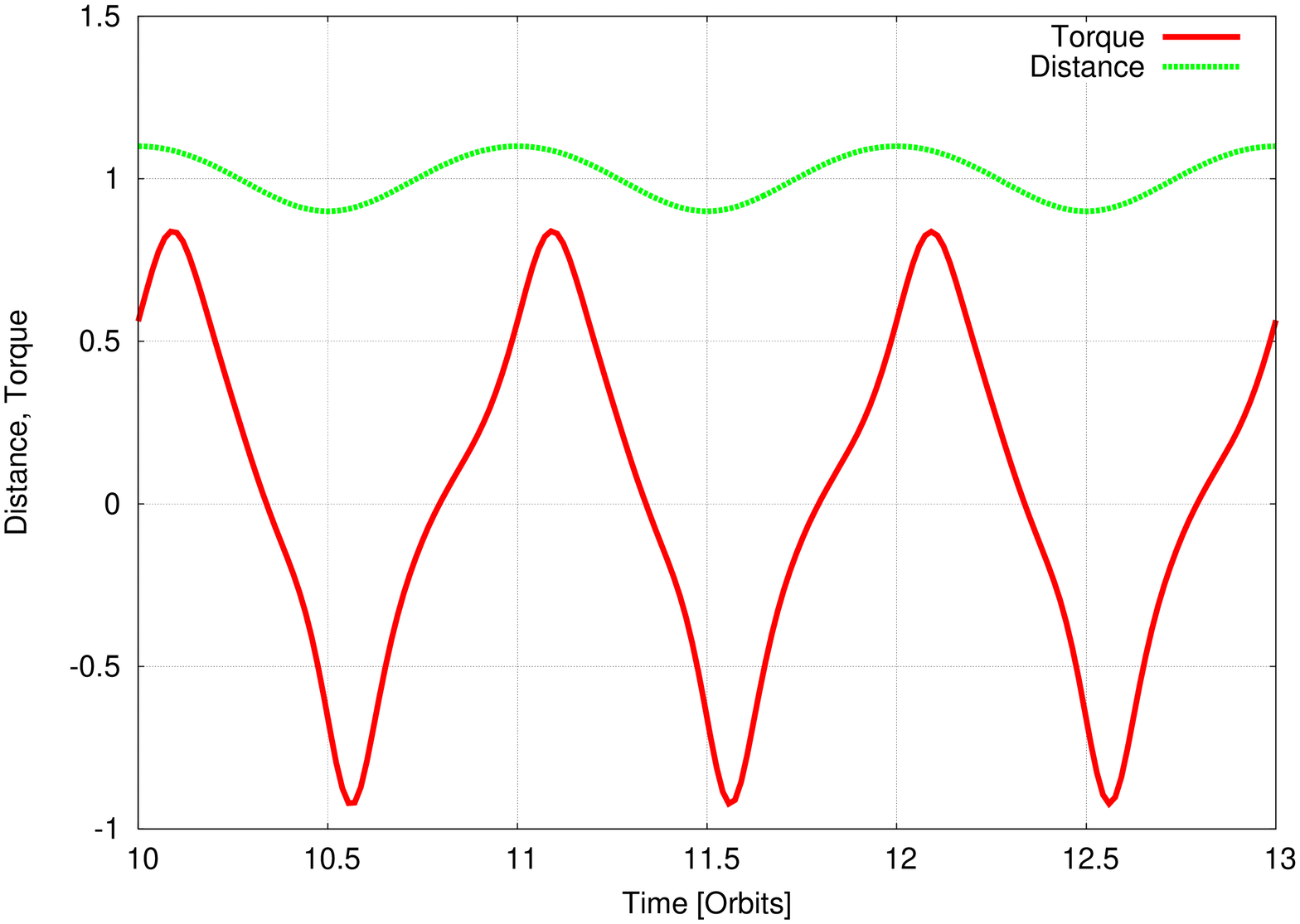}}}
\rotatebox{0}{
\resizebox{0.98\linewidth}{!}{%
\includegraphics{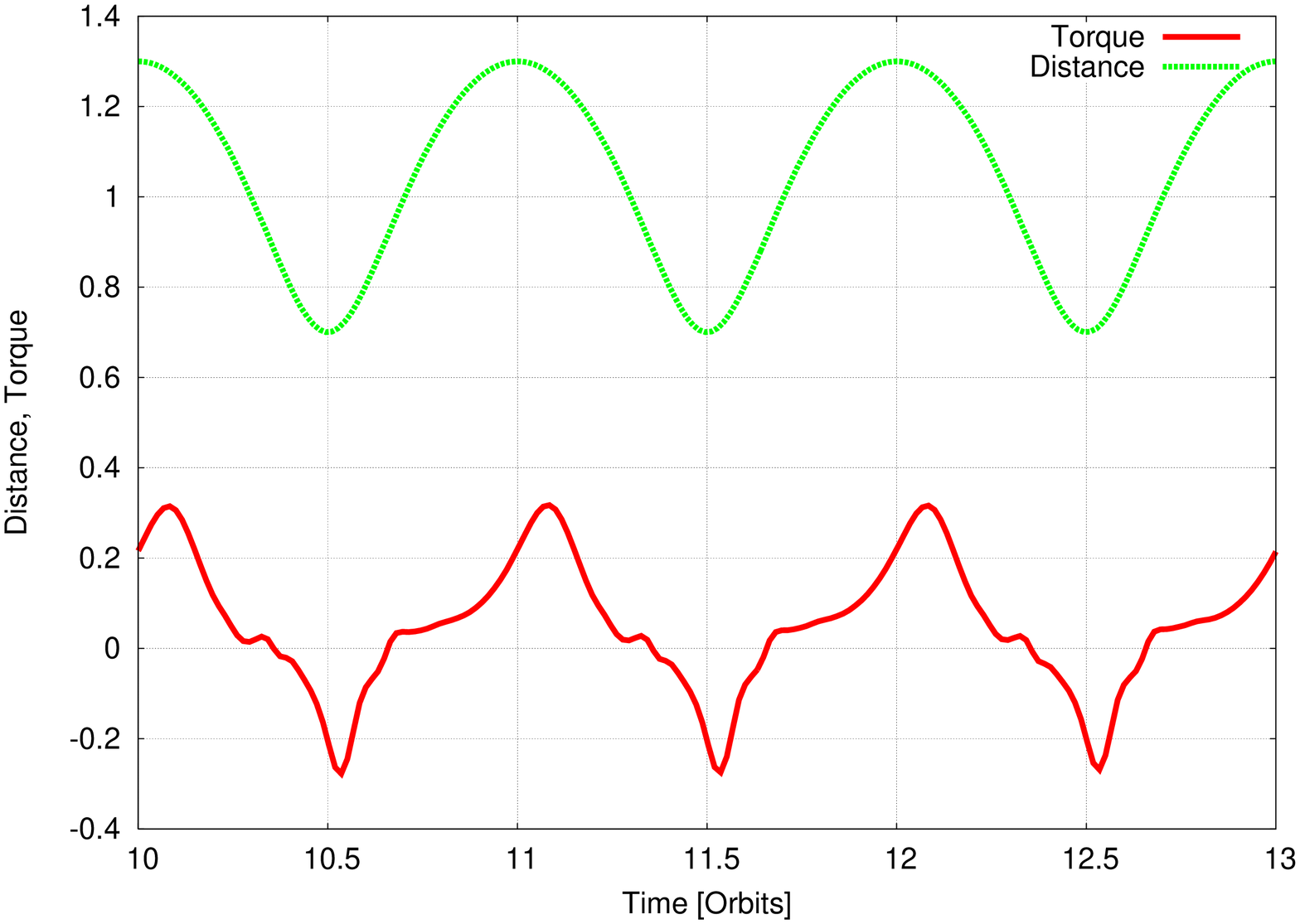}}}
\end{center}
  \caption{The change of the distance of the planet from the star and total disk torque acting on 
  the planet as a function of time, where the torque is rescaled appropriately.
  The planet is held fixed on its orbit.
  {\bf Top:} $e = 0.1$ and {\bf Bottom:} $e=0.3$.
 }
   \label{fig:tpos}
\end{figure}

This is exemplified in Fig.~\ref{fig:tpos} where we display
the variation of the torque (see Eq.~\ref{eq:torq})
acting on the planet and its radial distance form
the star as a function of time
for two different planetary eccentricities (top: $e = 0.1$, 
and bottom: $e = 0.3$).
The top plot ($e=0.1$) refers directly to  Fig.~\ref{fig:2dflow}.
Clearly in apoastron (strong inner spiral and leading high density feature)
the total total torque is positive,
while at periastron (strong outer spiral and lagging high density feature) 
the contribution is negative.
Additionally there appears to be a small phase shift between 
the distance and the torques.
Typically, for an embedded protoplanet the direction of migration 
and its magnitude
is attributed to the sign and value of the torque acting on it.
For the highly eccentric case ($e=0.3$) the 
average torque is clearly positive and one might expect outward migration.

However, this conclusion must be corrected for possible changes in the 
eccentricity of the planet. 
For a planet on an eccentric orbit its angular 
momentum $ L\subscr{p}$ is given by
\beq
      L\subscr{p}  =  m\subscr{p} \, \sqrt{ G M_* a} \, \sqrt{1 - e^2}
\eeq
and the  rate of change of the semi-major axis and eccentricity
can be obtained from 
\beq
\label{eq:ldot}
   \frac{\dot{L}\subscr{p}}{L\subscr{p}}
   = \frac{1}{2} \,  \frac{\dot{a}}{a} 
   - \frac{e^2}{1-e^2} \, \frac{\dot{e}}{e}
   \  =  \,  \frac{T\subscr{disk}}{L\subscr{p}}
\eeq
Here $T\subscr{disk}$ is the total torque exerted by the disk onto
the planet
\beq
\label{eq:torq}
    T\subscr{disk}  \, = \, \int_{Disk} \, \left. ( \vec{r}_p \times \vec{F}) \right|_z \, d f
\eeq
where $\vec{r}_p$ denotes the radius vector from the star to the planet,
$\vec{F} $ the (gravitational) force per unit area between the planet and a
disk element (at location $\vec{r}$ from the star), 
and $df$ the surface element.
Eq.~(\ref{eq:ldot}) implies that a positive torque may also result in eccentricity
damping rather than outward migration.
From our simulations we are able to disentangle the different
contributions.

\begin{figure}[ht]
\begin{center}
\rotatebox{0}{
\resizebox{0.98\linewidth}{!}{%
\includegraphics{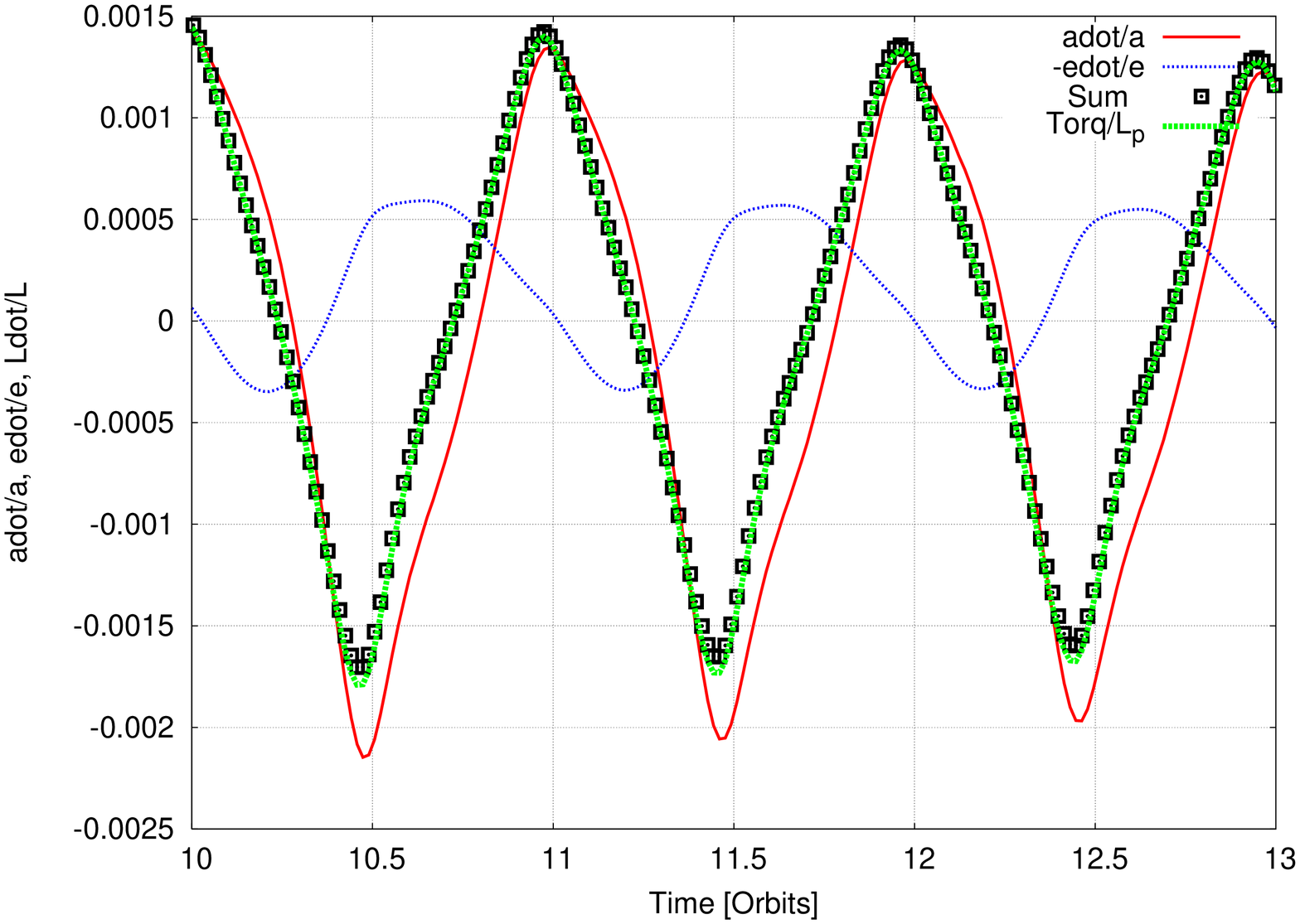}}}
\rotatebox{0}{
\resizebox{0.98\linewidth}{!}{%
\includegraphics{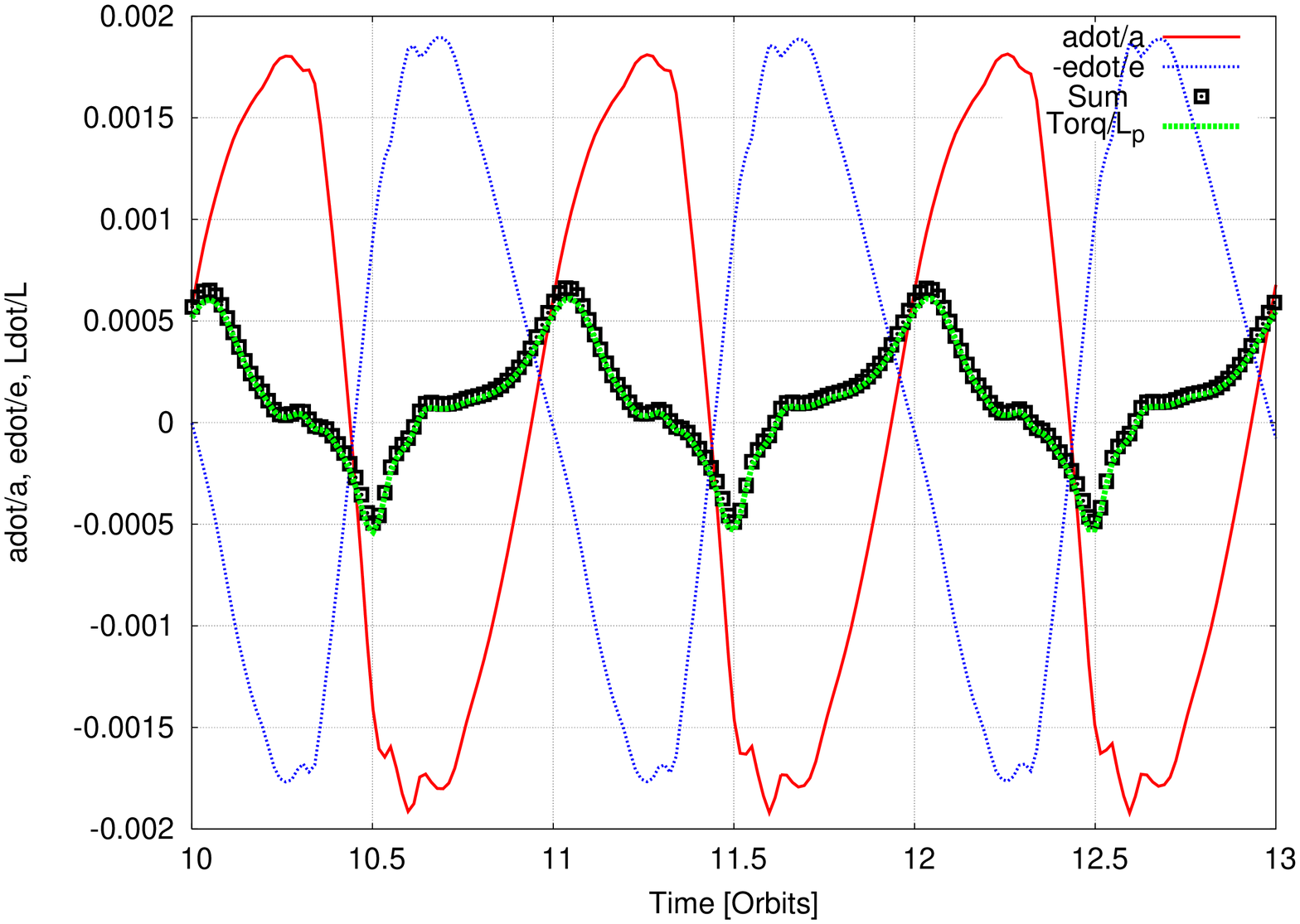}}}
\end{center}
  \caption{The time changes of the semi-major axis ($1/2 \dot{a}/a$, solid curve), the
  eccentricity ($ - e^2/(1-e^2) \, \dot{e}/e$, thin dashed line), and the normalised torque
  ($T\subscr{disk}/L\subscr{p}$) for an {\it evolving} eccentric planet in a 2D disk.
  The sum of the first two contributions is given by the symbols.
  {\bf Top:} $e_0 = 0.1$ and {\bf Bottom:} $e_0=0.3$.
 }
   \label{fig:aedot}
\end{figure}
  
The evolution of the semi-major axis and eccentricity changes are displayed
in Fig.~\ref{fig:aedot} for two different initial eccentricities in the 
time interval 10 to 13 orbits. After this time the flow has equilibrated and the
contribution of the individual terms in Eq.~(\ref{eq:ldot}) can be analysed.
In these calculations we have used a smoothed torque cutoff to reduce
the noise in the curves.
In both cases ($e_0 = 0.1$ and $e_0 = 0.3$) the sum of the two terms
for $\dot{a}$ and $\dot{e}$ in Eq.~(\ref{eq:ldot}) (thick dashed line) equals exactly 
the total disk torque as given by the squared symbols.
The periodic behaviour of all quantities is again clearly visible, they
all fluctuate around their zero value. 
Quite clearly, the contribution of the $\dot{e}$ term has comparable magnitude
to the  $\dot{a}$ value. Hence, a positive average total disk torque as for example in 
the $e=0.3$ case does not necessarily imply an outward migration of the planet,
since the total torque is
``shared'' between semi-major axis and eccentricity change (see Eq.~\ref{eq:ldot}).
A positive torque implies only that the angular momentum of the planet has to increase.
For an eccentric orbit this can be achieved in two ways,
either by increasing $a$ (outward migration) or by 
reducing $e$ (circularization). Indeed, for the semi-major axis
to increase the planet's total energy must increase. As described below,
we find that it in fact decreases even though the angular momentum
increases through the positive torque.

\subsection{Migration rate}
As noticed in the previous sections the migration rate 
depends on the eccentricity of the
embedded planet and can vary by about 
60\% (cf. Fig.~\ref{fig:migfunec}), an effect seen
in the 2D as well as in the 3D simulations. 
For small eccentricities, that are less than or equal to about 
twice the disk aspect ratio ($e \leq 2 H/r$),
we find an increase in the migration rate.
For larger values of $e$ the rate is reduced with 
respect to the circular case, but always directed inward.
To analyse this effect we have performed additional simulations 
in 2D with varying eccentricities.
To measure the values of the torque, the planet was held again on a 
fixed orbit.

\begin{figure}[ht]
\begin{center}
\rotatebox{0}{
\resizebox{0.98\linewidth}{!}{%
\includegraphics{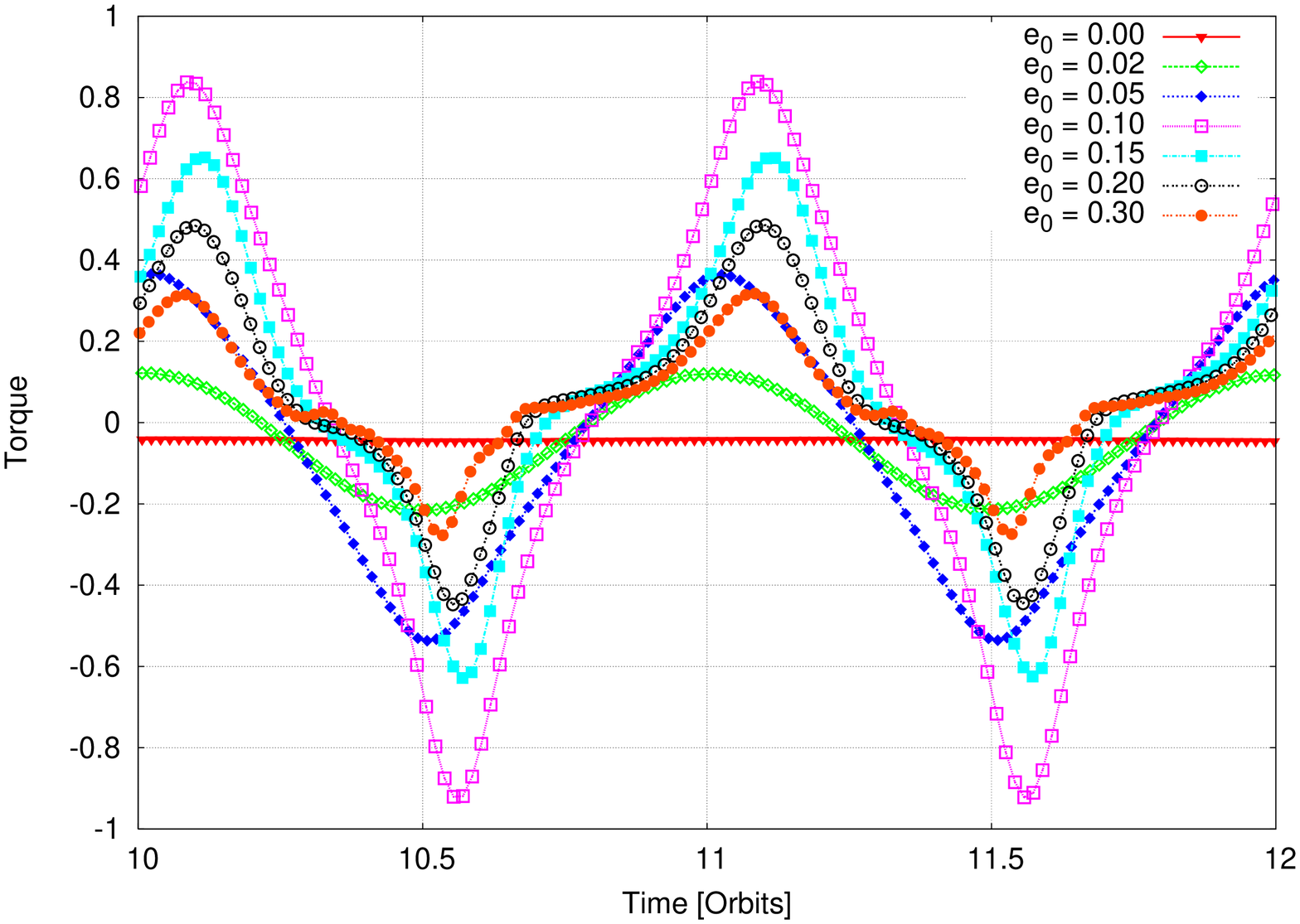}}}
\rotatebox{0}{
\resizebox{0.98\linewidth}{!}{%
\includegraphics{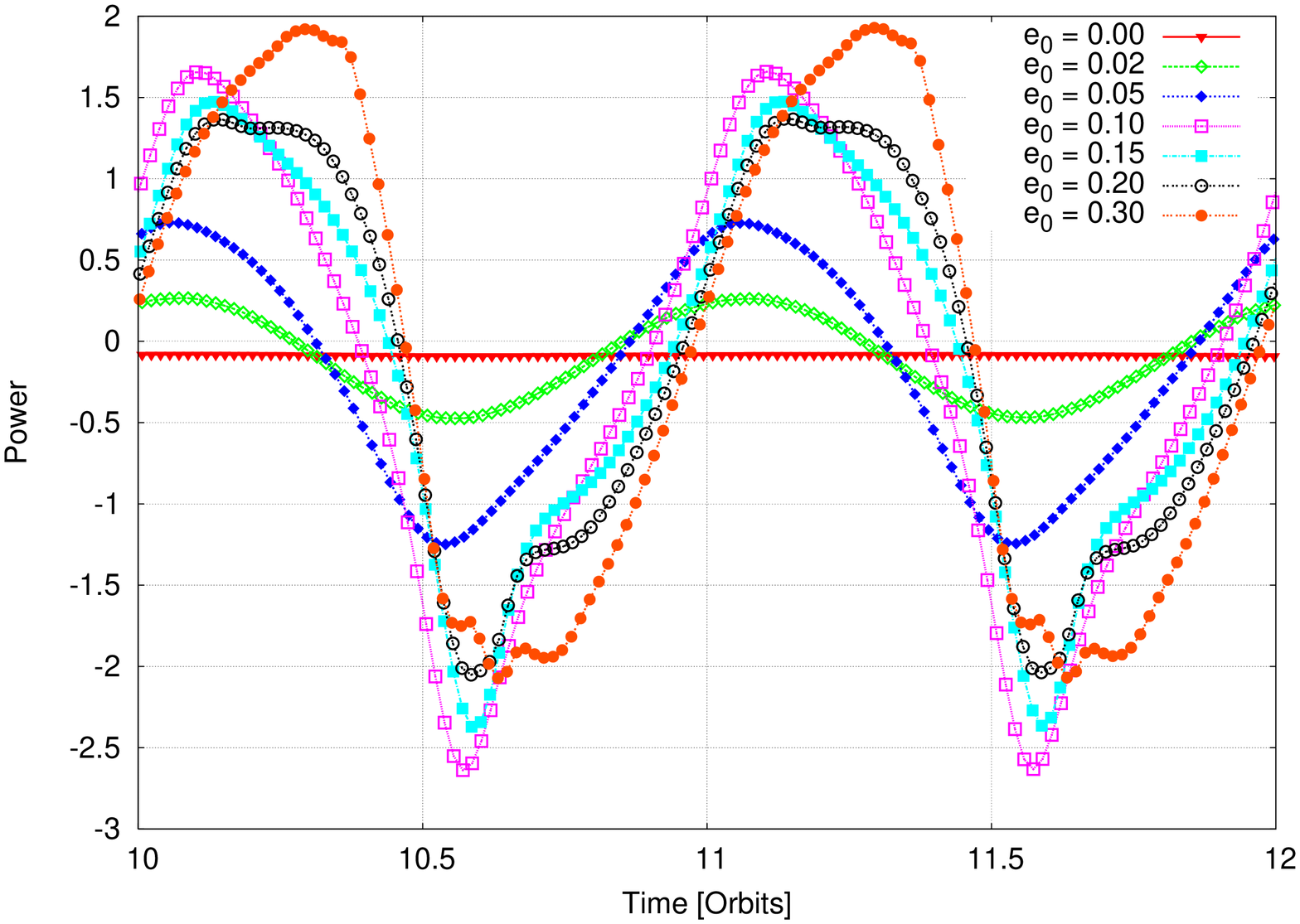}}}
\end{center}
  \caption{{\bf Top:} Normalised Torque ($T\subscr{disk}/{L\subscr{p}}$) vs.
   time for various eccentricities, where the orbit is not
   allowed to evolve. {\bf Bottom:} the corresponding energy loss
    ($P\subscr{disk}/{|E\subscr{p}|}$).
    The values have been amplified vertically by  
    the same amount as in Fig.~\ref{fig:tpos}. 
    }
   \label{fig:torq-p}
\end{figure}

In Fig.~\ref{fig:torq-p} we display the the evolution of the torque (top) and
mechanical power (bottom) as a function
of time over 2 orbital periods for different eccentricities.
Here the energy change per time (power) of the planet due to the work done
by the gravitational forces of the disk is given by
\beq
    P\subscr{disk}  \, = \, \int_{Disk} \, \vec{r}\subscr{p} \cdot \vec{F} \, d f
\eeq
The energy of the planet depends only on the semi-major axis $a$
and is given by
\beq
     E\subscr{p} \, = \, - \, \frac{1}{2} \, \frac{G M_* m\subscr{p}}{a}
\eeq
For the energy loss and semi-major axis change we then obtain
\beq
   \frac{\dot{E}\subscr{p}}{|E\subscr{p}|}
   =  \frac{\dot{a}}{a} 
   \  =  \, \frac{P\subscr{disk}}{|E\subscr{p}|}
\label{eq:edot}
\eeq
Restriction to {\it circular orbits} with $e=0$ yields
\beq
    \frac{\dot{E}\subscr{p}}{|E\subscr{p}|}   =  
     2 \,   \frac{\dot{L}\subscr{p}}{L\subscr{p}}
\eeq
or for the normalized torque and power
\beq
     \frac{P\subscr{disk}}{|E\subscr{p}|}  =   2  \, 
     \frac{T\subscr{disk}}{L\subscr{p}} 
\eeq
This relationship which can be directly verified from 
Fig.~\ref{fig:torq-p} where
the normalised torque and energy loss is displayed as a function of 
time for various eccentricities.
For low eccentricities ($e=0$: inverted triangles, $e=0.02$: open diamonds)
the shape of the curves are in good agreement, and 
only scaled by a factor of 2.
However, for eccentric non-circular motion this simple relation is not satisfied
anymore.
 
Already at very low eccentricities the variations of the power
exceed the mean value for $e = 0$ by a large margin.
For zero eccentricity the value for the power is about -0.09 
in the displayed units (solid curve with inverted triangles).
For $e=0.02$ (light dashed curve with open diamonds)
we find an amplitude of 0.37, and for 
$e=0.05$ (dark dashed curve with filled diamonds)
it has increased to 1.00 (bottom plot in  Fig.~\ref{fig:torq-p}). 
Hence, a small asymmetry in the power may lead to a substantial
change in the migration rate. 
From Fig.~\ref{fig:torq-p} it is clear that 
in particular during periapses ($t=10.5$ and $11.5$) the planets experiences
a large energy loss which reaches a maximum for an eccentricity around $e=0.10$. 
As a consequence the mean value (of $P_{disk}$)
has significantly dropped below the circular case,
leading to the enhanced inward migration found previously.
For larger $e$ the amplitude of the power variation
remains approximately at the same value but becomes more symmetric,
leading to a reduction in the migration rate.
As the orbit--integrated power remains negative for the high eccentricity
cases (i.e. $e \simeq 0.3$), the migration remains inward even
though the orbit integrated torque is positive. 

The variation of the torque for non-vanishing eccentricities 
also increases substantially above the
zero eccentricity value (Fig.~\ref{fig:torq-p} top panel).
Here, for larger values of $e$ above about $e=0.1$
the mean value becomes clearly positive. However, due to the
corresponding change in eccentricity and energy the migration is still
directed inwards (cf. Figs.~\ref{fig:tpos} and \ref{fig:aedot}).
 
\begin{figure}[ht]
\begin{center} 
\rotatebox{0}{
\resizebox{0.98\linewidth}{!}{%
\includegraphics{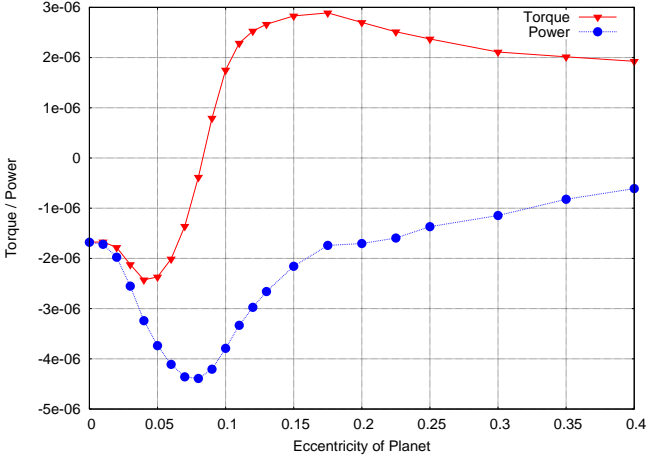}}}
\end{center}
  \caption{Dimensionless torque ($T\subscr{disk}$) and
   energy loss (power, $P\subscr{disk}$) experienced by the planet
   for different eccentricities. The values are obtained by keeping the
   planet on a fixed orbit and using time averages over 20 orbits
   (from $t=80$ to $100$). 
    }
   \label{fig:tpc1}
\end{figure}

To illuminate this effect from a different perspective we have performed a set 
of simulations where the planet was not allowed to move and remained on a fixed
orbit during the simulations. We used 22 different values for the eccentricity
ranging from $e=0$ to $e=0.40$ and ran all models up to a final time
of 100 orbits where the torques and the power acting on the planet have reached an
equilibrium on average. For the last 20 orbits (from $t=80$ to $100$) we 
calculate the time average of $T\subscr{disk}$ and $P\subscr{disk}$.
All these models are performed using the two-dimensional setup and utilize
a $256\times700$ grid with a logarithmic spacing in the radial direction.
Additionally, to increase performance the runs use the {\tt FARGO}-algorithm
for differentially rotating flows \citep{2000A&AS..141..165M}.
The results are displayed in Fig.~\ref{fig:tpc1} where the inverted triangles
refer to the torque and the solid dots to the power acting on the planet,
both given in dimensionless units.
Clearly, as already found through the previous analysis an increase in the
eccentricity leads to a non-monotonic behaviour of both $T\subscr{disk}$
and $P\subscr{disk}$. For circular motion both are identical and negative,
leading the the well known inward migration of the planet.
Upon increasing the eccentricity both quantities initially drop even more
and increase later again. The power remains negative for all eccentricities
leading to the inward migration which only slows down for larger $e$.
This leads to the behaviour of migration time scales as shown in
Fig.~\ref{fig:migfunec} above, which are shortest for $e \approx 0.08$.
The torque becomes positive for all $e > 0.08$ and reaches a maximum for
$e \approx 0.17$. 
 
\begin{figure}[ht]
\begin{center}
\rotatebox{0}{
\resizebox{0.98\linewidth}{!}{%
\includegraphics{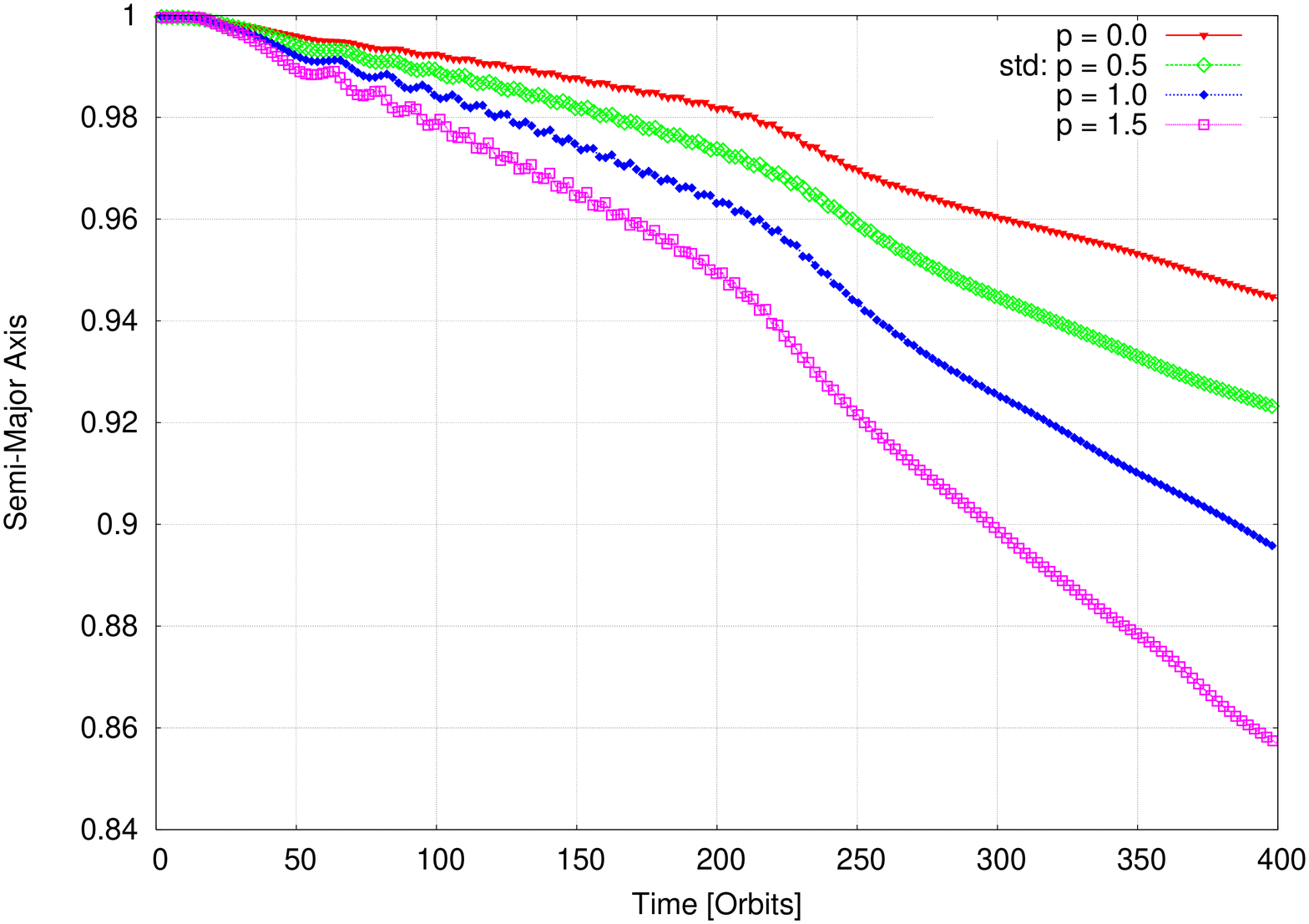}}}
\rotatebox{0}{
\resizebox{0.98\linewidth}{!}{%
\includegraphics{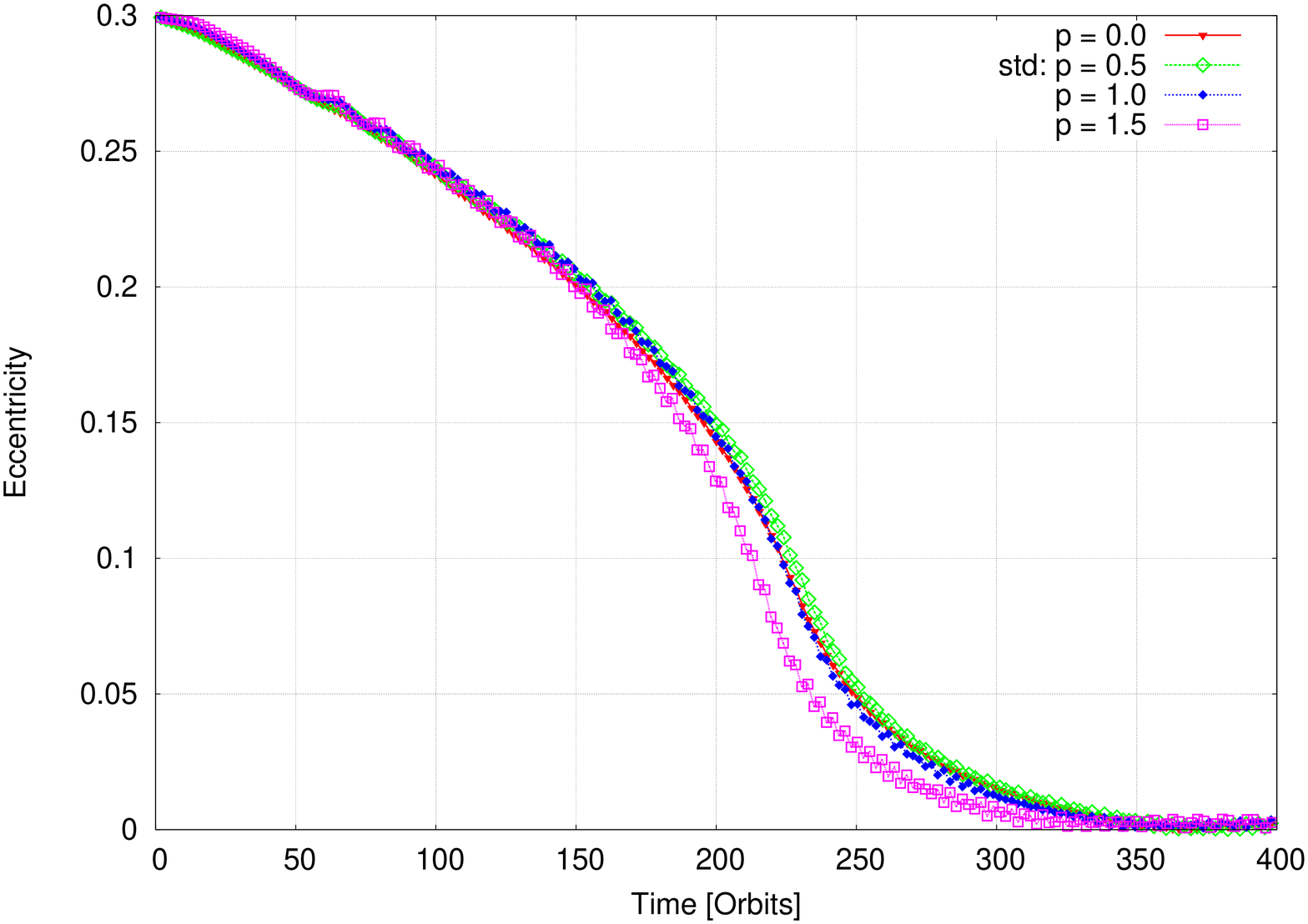}}}
\end{center}
  \caption{{\bf Top:} Time-change in semi-major axis for different surface density
   profils of the disk, given by $\Sigma(r) \propto r^{-p}$.
  {\bf Bottom:} Corresponding eccentricity evolution.
    }
   \label{fig:aesig}
\end{figure}

{\bf
\subsection{Changing the disk's density profile}
All the previous simulations have been peformed using a surface density
profile which varies as $\Sigma(r) \propto r^{-p}$, where we used $p=1/2$
which is the  equilibrium profile for a constant kinematic viscosity and
closed radial boundaries ($\dot{M}=0$). The corotation torques acting on the
planet and the resulting changes in migration speed depend (for circular orbits)
on the density slope of the disk \citep{2002ApJ...565.1257T, 2006ApJ...652..730M}.
To test the possible influence of the density slope on the orbital evolution of
eccentric (non-inclined) planets, we have varied the value of $p$ from
0 to 1.5 for our highest value of the eccentricity, $e=0.3$.
The results of our simulations are displayed in Fig.~\ref{fig:aesig}.
For an increasing  density slope the migration occurs monotonically at a faster rate,
in agreement
with standard linear estimates for the torques \citep{2002ApJ...565.1257T}.
The eccentricity evolution is much less affected, the largest change occurs only
for the steepest slope $p=1.5$. 
These results indicate that in the case of eccentric planets the importance of the
corotation torques are greatly reduced.
This is probably because during each orbit, the  planet moves a distance radially
that is larger than the width of the horseshoe region, where the corotation torque is
generated. Clearly it is expected that the corotation torque will weaken significantly
under these conditions.
In all four cases with different $p$ the average torque is clearly positive
while the power is negative. As in the previous $p=1/2$ case this leads to inward migration
and eccentricity damping. In particular our results concerning the positive torque
are in agreement with the earlier findings of \citet{2000MNRAS.315..823P} who used $p=1.5$,
but fixed the planetary orbit and did not measure the power.}

\section{Inclined orbits}
\label{sec:inclination}

In addition to the eccentric planetary orbits we now study the
additional degree of freedom provided by inclined orbits.
We begin by considering inclined, circular orbits,
before discussing the general case when both $e$ and 
$i$ are non-vanishing initially.
\subsection{Circular Orbit}
\begin{figure}[ht]
\begin{center}
\rotatebox{0}{
\resizebox{0.98\linewidth}{!}{%
\includegraphics{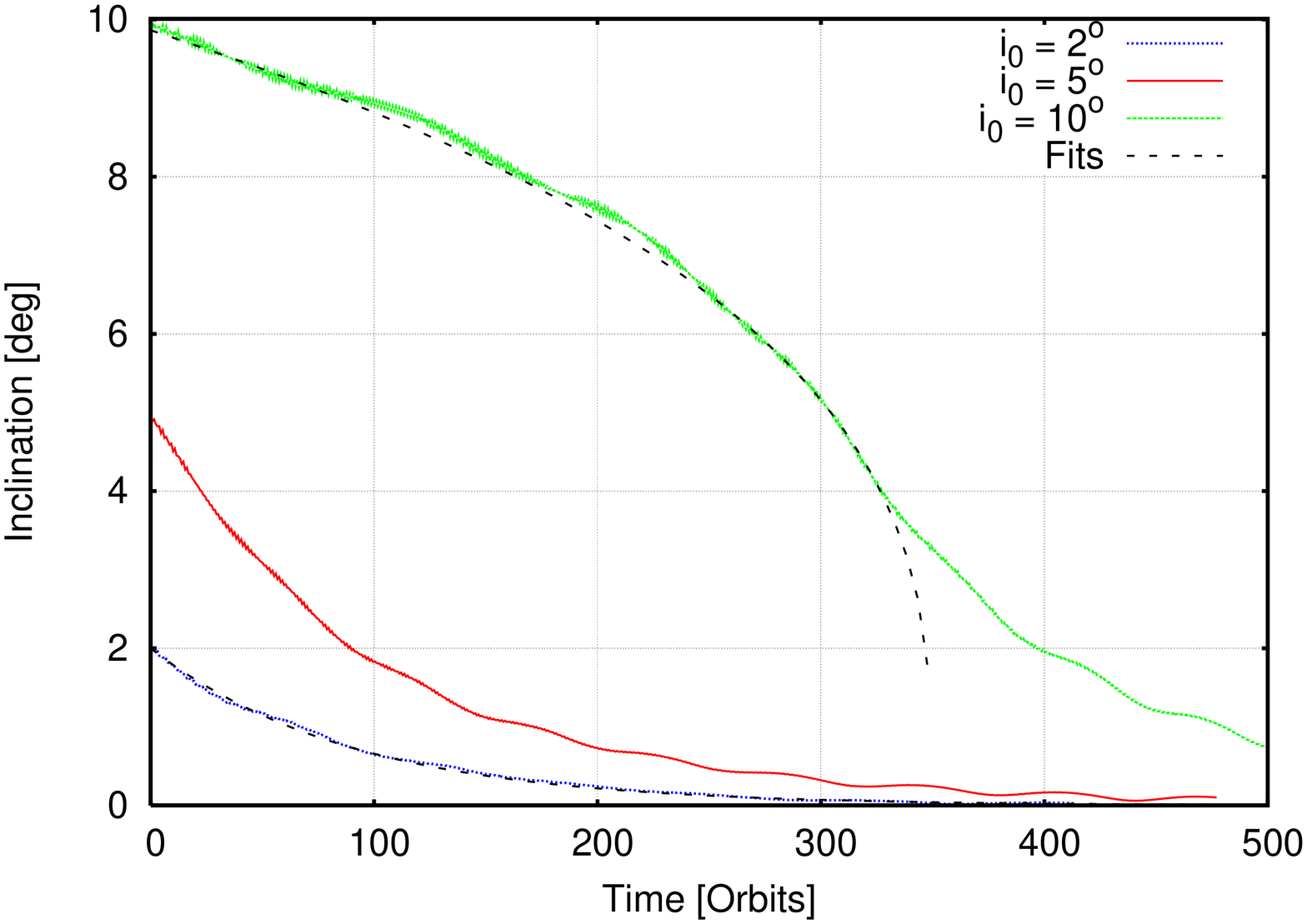}}}
\rotatebox{0}{
\resizebox{0.98\linewidth}{!}{%
\includegraphics{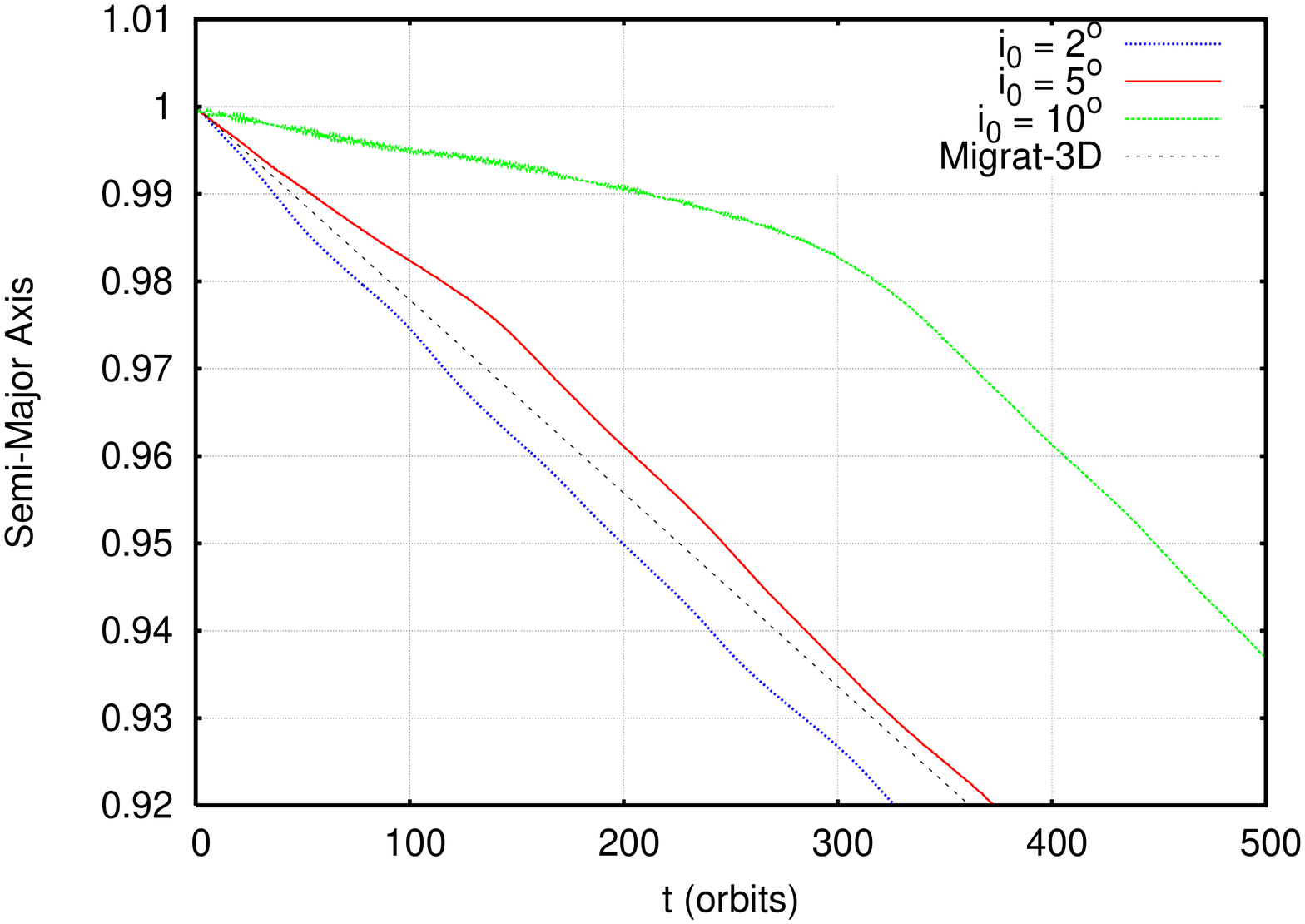}}}
\end{center}
  \caption{Orbital inclination (top panel) and semi-major axis (lower panel)
    as a function of time for different initial
    values of $i$ for a 20 Earth-mass planet on a {\it circular} orbit. 
    We find exponential decay of $i$ as long as $i \leq 6^\circ$; the thin dashed
   line superimposed on the lowest $i_0=2^\circ$ curve corresponds to an
   exponential with a damping timescale
    $\tau_{inc} =90$. For
    higher inclinations the damping timescale increases
    dramatically. A very good fit is found for a model with 
     $di/dt \propto i^{-2}$ (thin dashed line for $i_0=10^\circ$).}
   \label{fig:idamp}
\end{figure}

If we start a low mass planet on a circular but inclined orbit the orbital
inclination will be damped (cf. Fig.~\ref{fig:idamp}). For $i \simeq e$,
the time scale 
for inclination damping is somewhat longer than for the
eccentricity damping, but of the same order of magnitude.
Again we find two different regimes. For $i < H/r$ ($i$ in radians)
the planet remains inside the main body of the disk and the damping is exponential: 
$di/dt \propto i$. The
damping rate (averaged over one planetary orbit) for a small planet mass
and small inclinations is obtained through linear calculations by \citet{2004ApJ...602..388T} as
\begin{equation}
\frac{\overline{di / dt}}{i}  \, = \frac{1}{\tau_{inc}} \, = - \frac{0.544}{t_{wave}}
\end{equation}
with $t_{wave}$ defined in equation~(\ref{eq:twave}). For our
simulations this gives an inclination damping time scale  $\tau_{inc}$ of about 68 orbits. The two
lower curves in the upper panel of Fig.~\ref{fig:idamp} show the inclination damping
for small initial inclinations (2$^{\circ}$ and 5$^{\circ}$ respectively).
The fitted dashed line for $i_0=2^\circ$ corresponds to an exponential damping 
with a timescale of 90 orbits. This obtained $\tau_{inc}$ is larger than the linear estimate 
\citep{2004ApJ...602..388T} by the same factor as the eccentricity damping time $\tau_{ecc}$.

For higher inclinations we find a slower inclination damping rate as
can be seen in Fig.~\ref{fig:idamp} for $i_0 = 10^{\circ}$.
For our disk parameters, 
the damping rate departs from being exponential for values of $i$
in the interval 6$^{\circ} < i < 8^{\circ}$.
Just as for the damping of eccentricities, we therefore find that the linear 
calculations of \citet{2004ApJ...602..388T} provide
a good approximation for over 
twice their formal range of applicability, 
{\it i.e.\ } up to $i \sim 2 H/r$ - where here $i$ is measured in radians.
For $i \geq 8^\circ$
the damping rate strongly deviates from being exponential, 
and the best fit to the model with
the highest initial inclination of 10$^\circ$ is given by $di/dt
\propto i^{-2}$ (upper dashed line in Fig.~\ref{fig:idamp}).
Interestingly, this dependency is identical to the behaviour
of the eccentricity damping for large eccentricities.
However, here the reduced damping comes from the fact that for inclinations
significantly larger than $H/r$ the planet's contact with the disk is 
reduced, and its velocity relative to local disk gas as it crosses
the midplane is increased.

As with increasingly eccentric planar orbits, the migration rate is observed to 
decrease with increased inclination, as shown in Fig.~\ref{fig:idamp} (lower panel). The
effect of an increased inclination is weaker than for a comparable value of 
eccentricity: circular, inclined orbits lead only to weaker 
torques on the planet, and not the possibility of torque sign reversal.
Migration rates return smoothly to values close to the planar, circular rate
once the inclination drops beneath about 5$^\circ$.
\begin{figure}[ht]
\begin{center}
\rotatebox{0}{
\resizebox{0.98\linewidth}{!}{%
\includegraphics{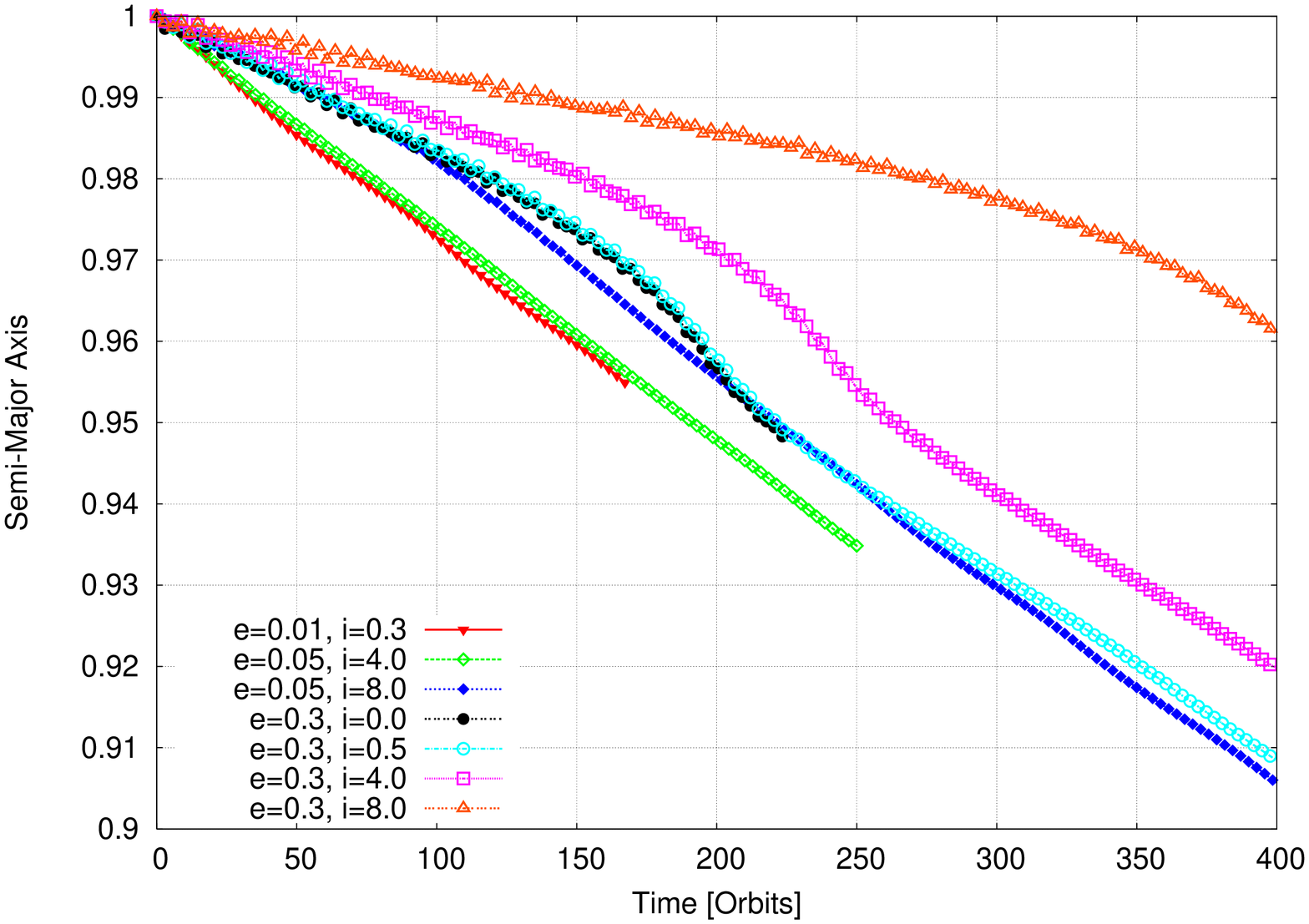}}}
\rotatebox{0}{
\resizebox{0.98\linewidth}{!}{%
\includegraphics{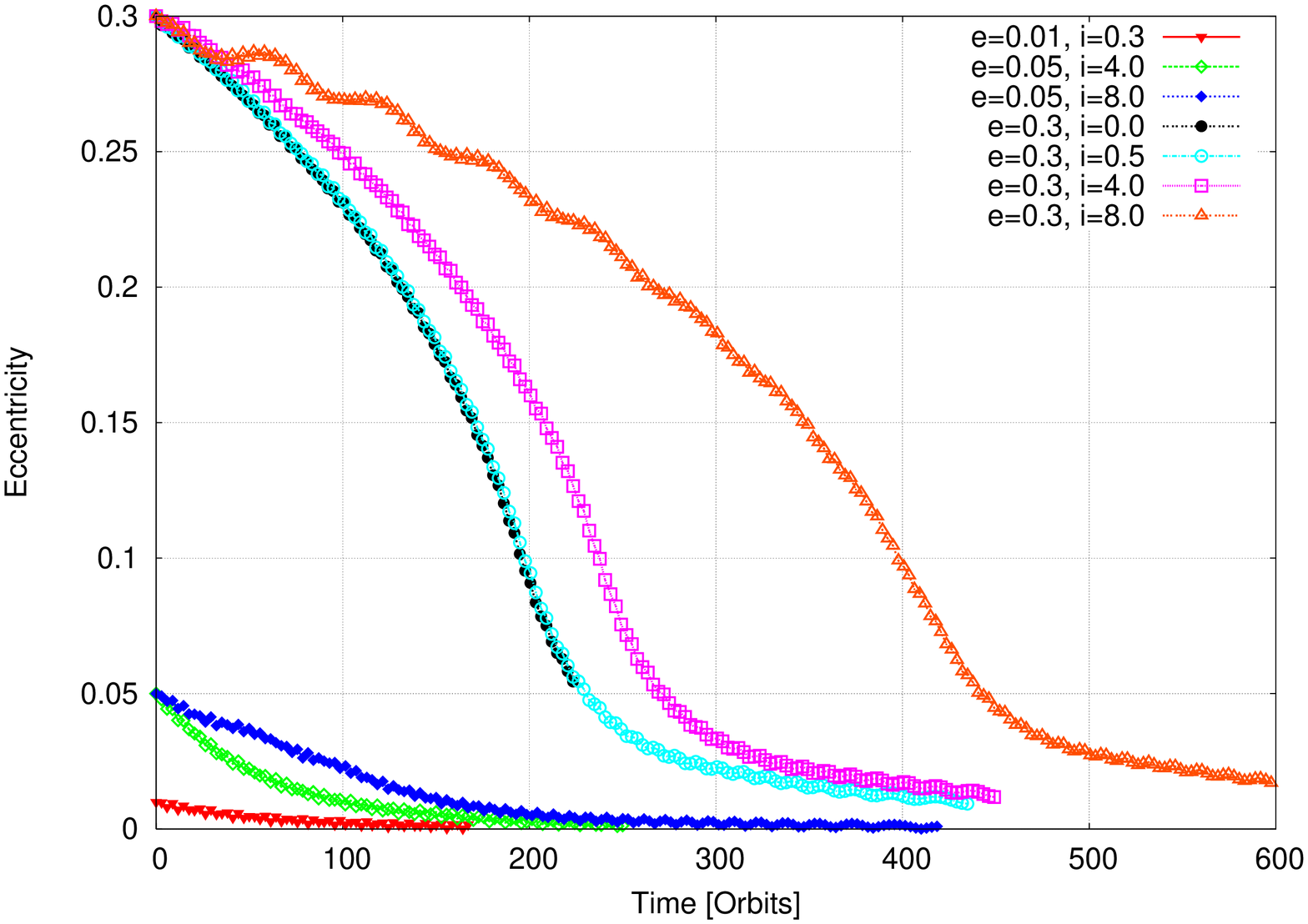}}}
\rotatebox{0}{
\resizebox{0.98\linewidth}{!}{%
\includegraphics{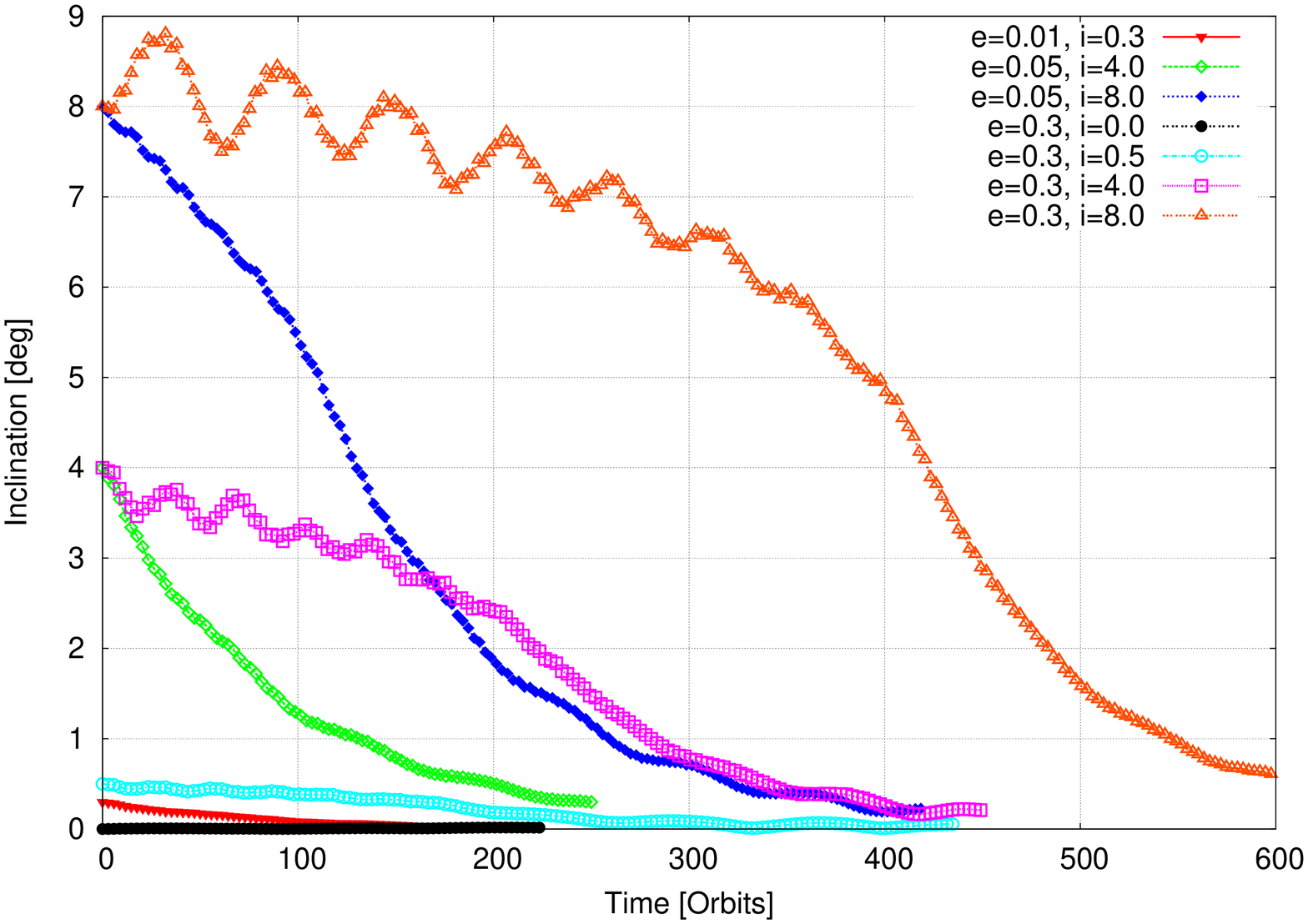}}}
\end{center}
  \caption{Semi-major axis, eccentricity and inclination as a 
   function of time for a 20 $m_E$ planet with different initial
   eccentricities and inclinations.
   In the top panel the time coordinate has been stretched to show
   effects more clearly.
   }
   \label{fig:eccmix}
\end{figure}
\subsection{Non-circular Orbits}
Finally we investigate several models in which both the initial
eccentricity and the initial inclination are nonzero with $0 < e_0 <
0.30$ and $0 < i_0 < 8^{\circ}$. Each system was initialised 
with the planet at apocentre above the midplane, with the
longitude of  pericentre $\omega=\pi/2$.
Our results are summarised in Fig.~\ref{fig:eccmix}.

\subsubsection{Small $e$ and small $i$}
We begin by considering the evolution of orbits for which
the planets are initiated to have small values of eccentricity
and inclination (i.e. $e_0 \lesssim H/R$ and $i_0 \lesssim H/R$,
where $i_0$ is measured in radians).

Examining the top panel of Fig.~\ref{fig:eccmix} we can see the
migration behaviour for runs with 
($e_0=0.01$ and $i_0=0.3^\circ$) and ($e_0=0.05$ and $i_0=4^{\circ}$).
It is clear that for small values of both $e_0$ and $i_0$
the migration behaves as if the orbit was essentially uninclined.
Small values of the inclination such that $i \le H/R$, with $i$ measured in
radians, cause migration to be slowed, but only very slightly.

Examining the middle panel of Fig.~\ref{fig:eccmix}, we can see that
small values of $e_0$ and $i_0$ lead to exponential decay of
the planet eccentricity, with the rate of damping only slightly increased
by the small inclination. Examining Fig.~\ref{fig:e3dx}
we see that the time taken for the $e_0=0.1$ case to halve its
eccentricity is $\simeq 35$ orbits, and the time required
for the $e_0=0.05$, $i_0=4^{\circ}$ case shown in Fig.~\ref{fig:eccmix}
to also halve its eccentricity is approximately 40 orbits. 

The lowest panel in Fig.~\ref{fig:eccmix} shows the inclination
evolution. In common with the eccentricity damping,
we find that the inclination damping for an orbit with small
initial values of $e$ and $i$ occurs exponentially on a time scale 
very similar to the case when the orbit is circular but inclined.

\subsubsection{Small $i$ and large $e$}
We now consider the orbital evolution when the inclination is
small ($i \lesssim H/R$) and the eccentricity is large ($e  > H/R$).
This is represented by the runs shown in Fig.~\ref{fig:eccmix}
with ($e_0=0.3$ and $i_0=0.5^\circ$) and ($e_0=0.3$ and $i_0=4^\circ$).
Also shown for reference is a case with ($e_0=0.3$ and $i_0=0$).
The migration is shown in the top panel. 
We have already shown in previous sections that an uninclined, high eccentricity
orbit undergoes slower inward migration than a circular orbit does.
We see that increasing the inclination increases the migration time further.
A very small inclination ($i=0.5^\circ$) has very little effect,
but an inclination of $i=4^\circ$ begins to make a noticeable
difference, decreasing the migration rate by $\simeq 25\%$.

The eccentricity evolution of these runs is shown in the middle
panel of Fig.~\ref{fig:eccmix}. The discussion presented
in previous sections showed that an uninclined, highly eccentric
orbit shows an eccentricity decay rate $de/dt \propto e^{-2}$.
We can see that a very small inclination ($i=0.5^\circ$)
has essentially no effect, but an inclination of $i=4^\circ$
extends the damping time by about 25\%, while preserving the shape
of the curve of $e$ versus time. 

The inclination evolution is shown in the lowest panel.
If we consider the evolution of the runs with $i_0=4^\circ$,
then we see a fairly dramatic change in the inclination damping
once $e_0 >> H/R$. For small values of $e_0$ we see the usual
exponential decline of $i$. When $e_0$ is large, however, the inclination
damping rate is small (even for small inclination)
and almost constant with superimposed oscillations until the eccentricity falls below
$e \simeq 0.1$, after which it recovers the exponential decay.
For example, the $i_0=4^\circ$ and $e_0=0.3$ case shows an 
inclination decay time scale $i_0/(di/dt)\sim 700$ orbits initially.
Once the eccentricity has damped to $e \le 0.1$ after approximately
200 orbits, the inclination declines exponentially on an e-folding
timescale of $\simeq 90$ orbits.
The period of the oscillations seen in the inclination corresponds approximately to the
precession period of the nodal line of the planet. 

\subsubsection{Large $i$ and small $e$}
We now consider the evolution when the inclination $i$ is
large and the eccentricity is small, which is represented in
Fig.~\ref{fig:eccmix} by the run with ($i_0=8^\circ$ and $e_0=0.05$).
The migration is shown in the top panel, and we can see that
compared to the run with ($i_0=4^\circ$ and $e_0=0.05$)
the migration rate for the more inclined planet is slower by about 40\%,
such that a doubling of the initial inclination leads to a 
migration rate that is almost halved. After just over 220 orbits
the migration rate approaches that for uninclined orbits
as the inclination has decreased sufficiently after this time.

The eccentricity evolution for this run is shown in the middle
panel of Fig.~\ref{fig:eccmix}.
Comparing this run with the one with ($i_0=4^\circ$ and $e_0=0.05$)
we see that having a large inclination substantially increases
the damping time. Close to the beginning of the simulation
the low inclination run has its eccentricity reduced by 
about 50\% within $\simeq 40$ orbits, whereas the more inclined run
shows only a 20\% change in eccentricity after this time.
The damping of eccentricity also shows a strong deviation from
exponential decay for the high inclination run, with the
eccentricity decay rate being almost constant until the
inclination has damped down to $\simeq 4^\circ$. Once the inclination
has declined down to this value the eccentricity decay becomes
exponential.

We have already discussed the inclination evolution for
circular orbits, and have shown that for large values of
$i$ the inclination damping rate is no longer exponential,
but scales as $di/dt \propto i^{-2}$.
The addition of a small eccentricity ($e_0=0.05$) barely changes the
decay rate for large inclinations, as shown in the lowest
panel of Fig.~\ref{fig:eccmix}. The effect of this small eccentricity
is simply to lengthen the damping time by about $10\%$.

\subsubsection{Large $i$ and large $e$}
We now consider the evolution of orbits with both large inclination and
eccentricity. The run with ($e_0=0.3$ and $i=8^{\circ}$), plotted
in Fig.~\ref{fig:eccmix} illustrates this case. The migration rate,
shown in the top panel, is seen to be reduced substantially
compared to both the cases with ($e_0=0.3$ and $i_0=4^{\circ}$),
and ($e_0=0.05$ and $i_0=8^{\circ}$). An increase of inclination
from $4^{\circ}$ to $8^{\circ}$, with eccentricity $e=0.3$
leads to a $\simeq 50\%$ reduction in the initial migration rate. 
However, we note that the rapid damping of inclination and
eccentricity means that this reduction in migration
rate is short lived, and can at most only increase
the total migration time through the disk by $< 10\%$
in the absence of a mechanism to maintain $e$ and $i$ at
large values \citep{2006A&A...450..833C}.

The eccentricity evolution shown in the middle panel
demonstrates that when $e_0=0.3$,
increasing the inclination from 
$4^\circ$ to $8^\circ$ causes a significant lengthening
of the initial eccentricity damping time (by about a factor of 2).
As for the other high eccentricity/inclination runs, the damping
is no longer exponential while $e > 0.1$ or $i> 4^{\circ}$,
but becomes so once the eccentricity and inclination fall
beneath these values.

The inclination damping shows similar behaviour to the
eccentricity damping for this large $e$ and $i$ run,
although the dependence of $di/dt$ on $i$ is slightly weaker
than that seen for $de/dt$. Increasing the inclination from
$i=4^\circ$ to $i=8^\circ$ with $e_0=0.3$ decreases
the initial inclination damping rate by about 40\%,
with little change to the overall shape of the curve.
We see that both the eccentricity and inclination decay
on similar time scales (within about 300 orbits for $e$ and 350 orbits for $i$),
and once they have reached small values they under go exponential
decay toward zero. 

\section{Conclusions}
\label{sec:conclusion}

We have performed fully non-linear 2- and 3-dimensional hydrodynamical simulations of a planet 
embedded in a protoplanetary disk. The planet was allowed to change its 
orbit due to the torques from the disk material. We
investigated the orbital evolution of the planet in the linear low mass regime 
(Type I migration) for which analytical studies of the eccentricity and inclination
damping rates are known. 

Concerning the damping of eccentricities and
inclinations, for small values $e < 0.1 $ and $i < 5^\circ$ we find very  
good agreement 
between our non-linear results and existing linear theory \citep{2004ApJ...602..388T}.
The damping occurs for both, $i$ and $e$, exponentially
with a time scale $\tau \propto (H/r)^2 \tau_{mig}$. 
The quantitative agreement of our numerical results with the estimates of the
linear calculations for the eccentricity and inclination damping are at the 20-30\% level,
which is very encouraging considering that one always has to use both gravitational
smoothing and torque--cutoff within the planet Hill sphere in numerical simulations. 
\citet{2000MNRAS.315..823P} have shown through linear calculations that
the absolute magnitude of the torque depends on the value of the gravitational 
smoothing $\epsilon$, and the slightly longer damping times that we find in
3D simulations is probably an indication that softening is playing a role. In addition,
\citet{2006ApJ...652..730M} have suggested that departures from the linear results
may occur for 20 Earth mass planets that we have considered, so we should probably
not expect perfect agreement with the results of \citet{2004ApJ...602..388T}.

For our adopted $H/r =0.05$, the 
critical values for $e$ and $i$ (in radians) lie approximately at $2 H/r$.
Above those critical values the time behaviour changes for both,
eccentricity and inclination.
In both cases we find a damping following $d y/dt \propto y^{-2}$ with $y \in \{ e, i\}$,
which (for the eccentricity) has been suggested recently by \citet{2000MNRAS.315..823P}.
For the inclination we attribute this behaviour to the fact that on highly inclined
orbits the planet loses contact with the disk and experiences only reduced damping,
while for the eccentricity it is has been attributed to the varying planetary velocity
with respect to the disk material \citep{2000MNRAS.315..823P}.

For the migration rates as a function of eccentricity we find some
surprising results: For low eccentricities in the exponential damping regime
we find faster migration than in the case of zero eccentricity
- an increase of up to 60\% is observed.
Through a detailed analysis of torque and energy loss balance during the 
planetary orbit we attribute this increased migration rate
to an increase in the energy loss
during periapse of the planet which reaches a maximum at $e \approx 0.1$.
For larger eccentricities the migration rate is reduced significantly below
the circular case but is still directed inward even for $e =0.3$, even though the
average torques are clearly positive. Migration can still be inward in this case
because the torque largely goes into damping of the eccentricity while the
planet continues to lose energy on average. If a situation were to arise where
this energy loss could be compensated for, for example by gravitational interaction with
surrounding planets, then in principle the positive disc torque could
drive outward migration. Simulations by \citet{2006A&A...450..833C} have explored
this possibility, but find that eventually the eccentricity of all planets
damps and inward migration of the planetary swarm occurs.
\begin{figure}[ht]
\begin{center}
\rotatebox{0}{
\resizebox{0.98\linewidth}{!}{%
\includegraphics{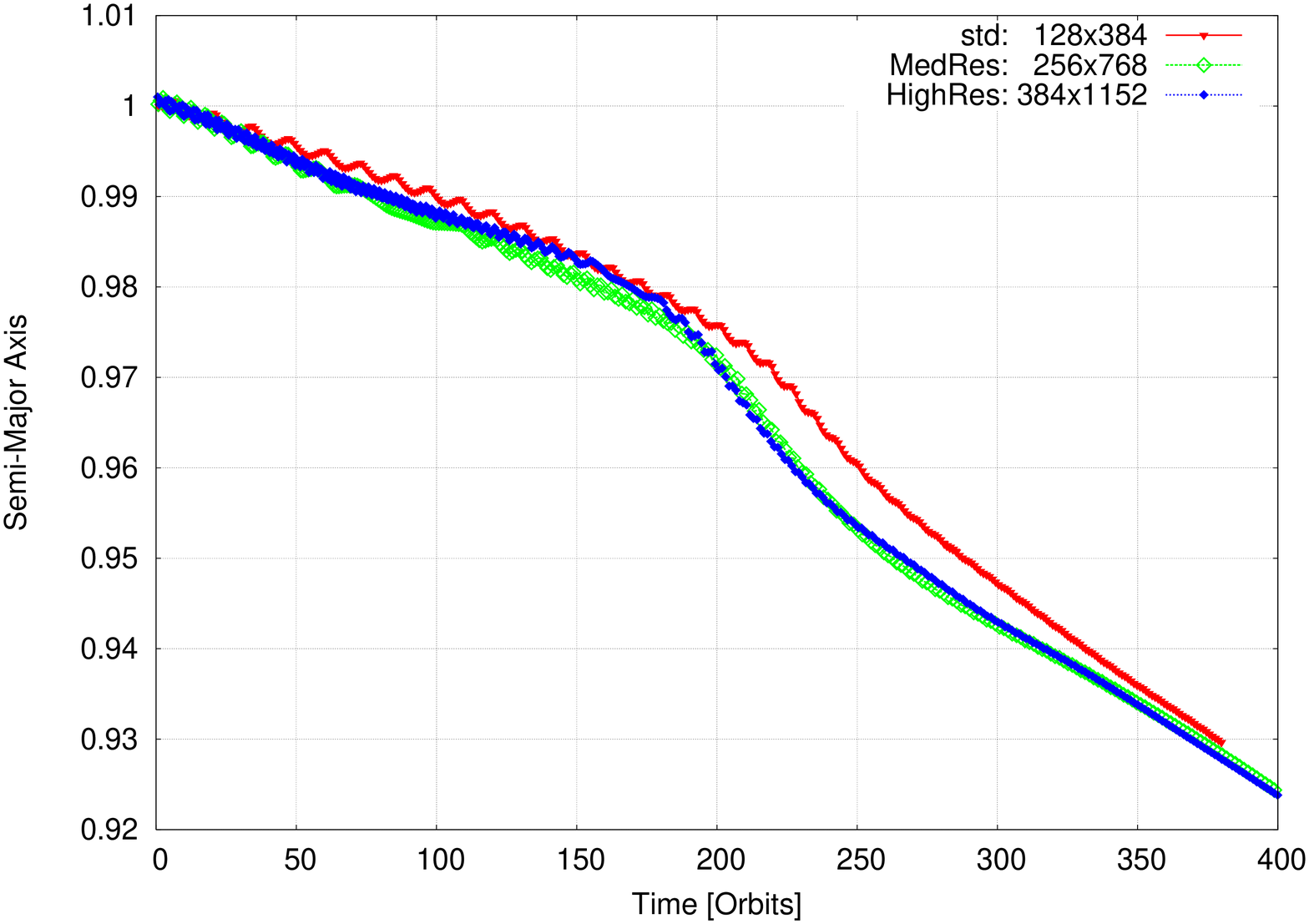}}}
\rotatebox{0}{
\resizebox{0.98\linewidth}{!}{%
\includegraphics{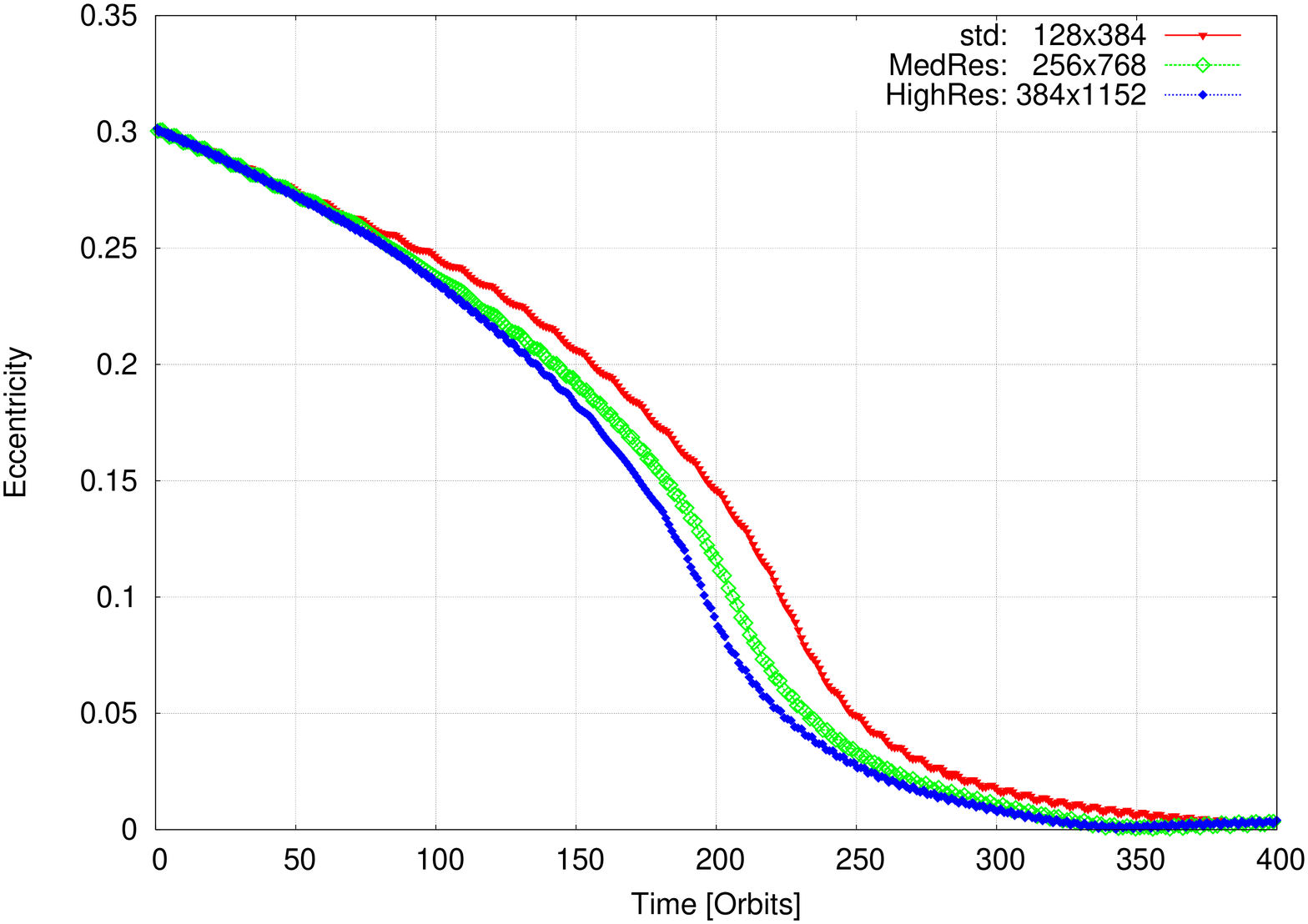}}}
\end{center}
  \caption{Semi-major axis and eccentricity as a
   function of time for a two-dimensional disk with a
  20 $m_E$ planet and an initial eccentricity
   $e_0=0.3$ for different grid resolutions.
   }
   \label{fig:aeres2d}
\end{figure}

We compare our 3D results in detail with corresponding 2D simulations in the
case of vanishing inclinations.
For coplanar orbits and small eccentricities the results of the 2D and 3D simulations
are in excellent agreement with respect to the migration rate and the eccentricity
damping time scale.
The increased migration for small non-zero
eccentricities is also reproduced. 
For eccentricities $e > 0.25$, agreement between the 2D and 3D is poorer.
While the qualitative behaviour is still very similar, we find
differences in the time scales.
For example, the eccentricity damping rate between 2D and 3D 
differs by 10-20\% for the $e_0 =0.3$ case.
As the planet samples a large range of cell sizes and the interaction 
becomes more sensitive to the smoothing length selected. This issue can 
be solved if detailed knowledge of the problem under examination is known
beforehand, but with the increased availability of high-performance and
parallel computing facilities, it is questionable as to whether the results
are worth the investment.
Hence, the evolution of embedded planets can be studied in 2D, for 
eccentricities of up to several scaleheights, if the right smoothing length
is chosen for the potential. For larger values of eccentricity, $e$-folding 
times are typically overestimated.

For the combined general case, non-zero eccentricity and inclined orbits, we
find that provided the eccentricity or inclination remains small, the previous
formulae remain a good approximation of a planet's behaviour. Inclination
produces a weaker departure from the circular, planar case than the same 
value of eccentricity, and non-linear effects set in sooner for lower $e$
when both parameters are non-zero. The 
migration rate may be moderately 
(factors of $\sim$ 2 -- 3) reduced if both $e$ and $i$
are towards the upper limits of this range. Once both parameters enter the 
non-linear regime however, cross-terms become significant in both the damping
and migration, and the previous correspondences ($de/dt \propto e^{-2}, 
di/dt \propto i^{-2}$) are lost. Secular effects from the disk can lead to 
moderate, short-term oscillations in $e$ and $i$.

Our results indicate that the interaction of a
small mass planet (here 20 $m_E$) with
its protoplanetary disk always leads to a rapid 
damping of its eccentricity and inclination.
Additionally, the main effects and timescales are 
captured very well by a two-dimensional
approach which greatly simplifies the computational effort.
Only if additional perturbers, such as other 
planets or a stellar binary companion,
are present may the eccentricity/inclination of the 
planet and the disk be excited.
\begin{acknowledgements}

Very fruitful discussions with Fred\'eric Masset are gratefully acknowledged.
The work was sponsored by the EC-RTN Network {\it The Origin of
Planetary Systems} under grant HPRN-CT-2002-00308.
Some of the computations presented in this work were
performed on the U.K. Astrophysical Fluids Facility (UKAFF),
and on the QMUL High Performance Computing Facility funded
under the SRIF initiative.

\end{acknowledgements}

\begin{figure}[ht]
\begin{center}
\rotatebox{0}{
\resizebox{0.95\linewidth}{!}{%
\includegraphics{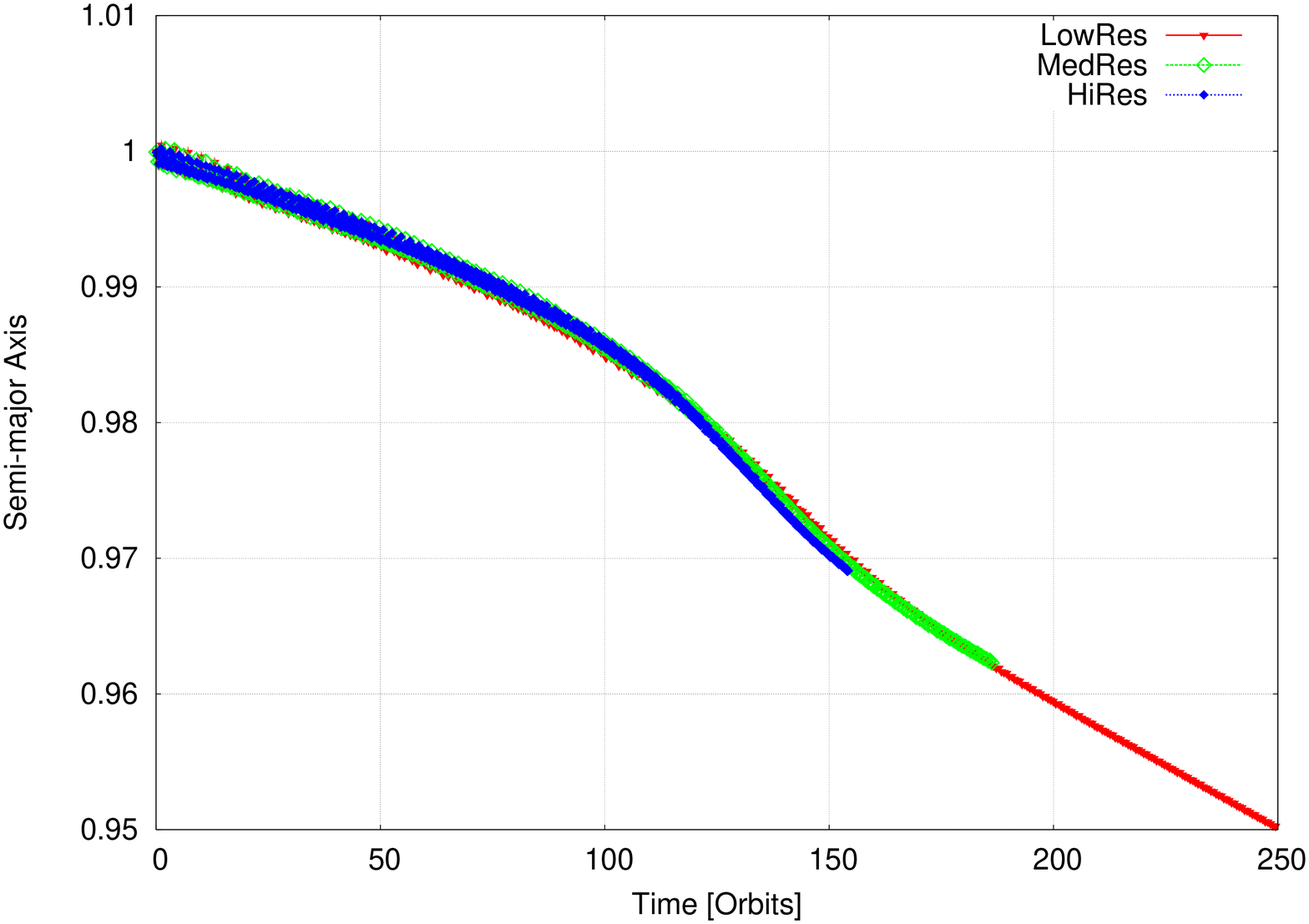}}}
\rotatebox{0}{
\resizebox{0.95\linewidth}{!}{%
\includegraphics{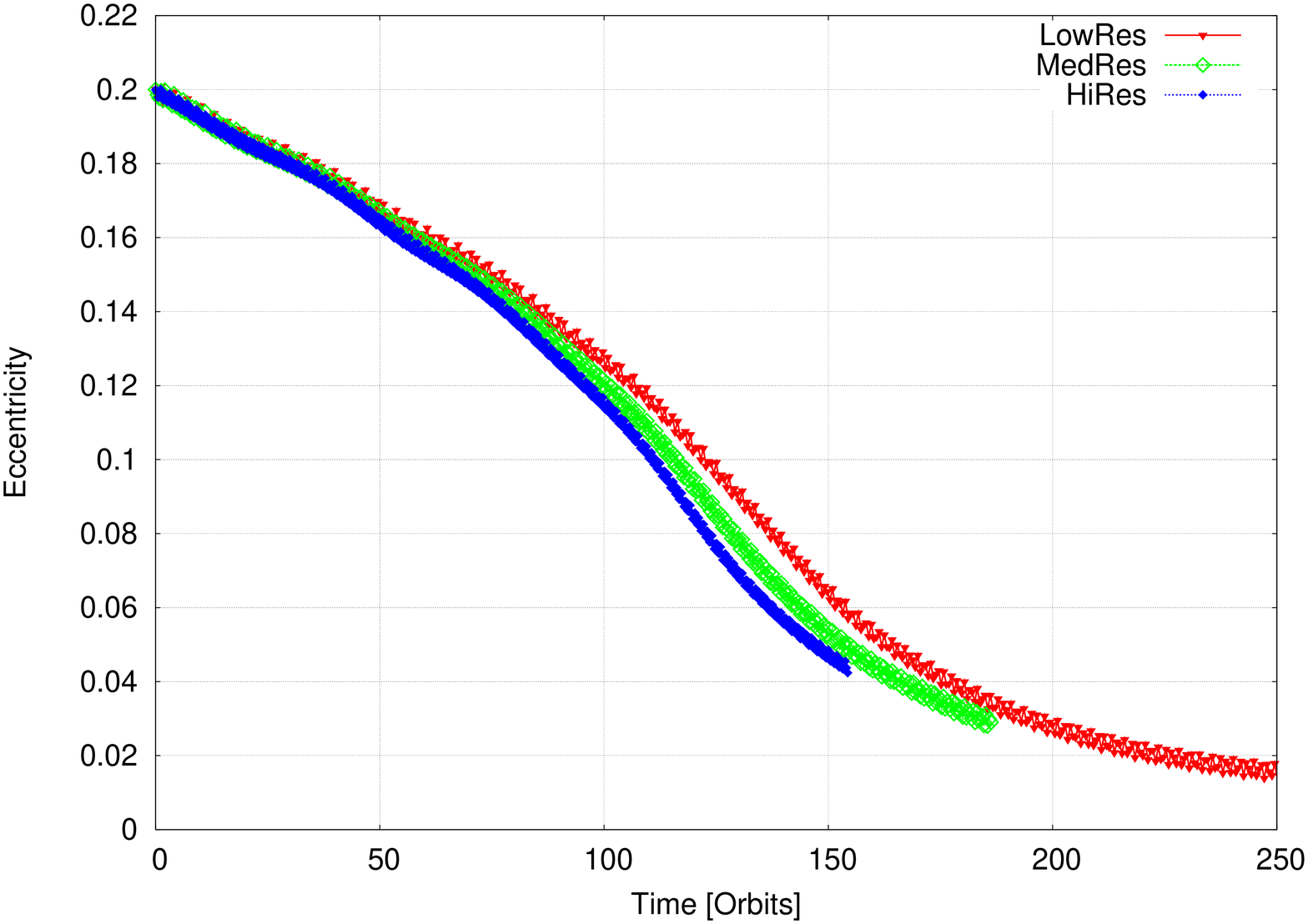}}}
\rotatebox{0}{
\resizebox{0.95\linewidth}{!}{%
\includegraphics{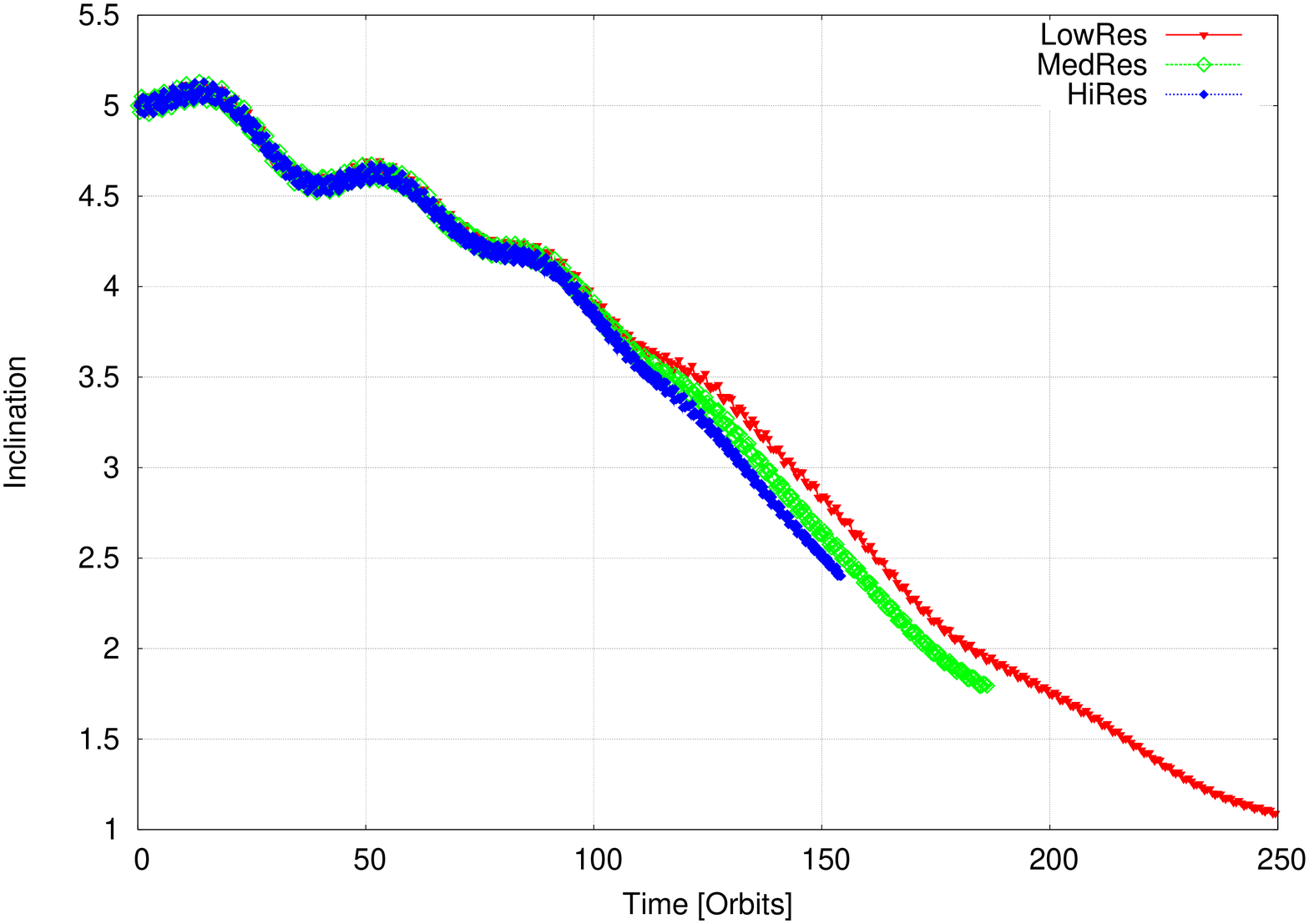}}}
\end{center}
  \caption{Semi-major axis, eccentricity and inclination as a
   function of time for a 20 $m_E$ planet in a three-dimensional disk,
   for different grid resolutions (see text).
   }
   \label{fig:aeires3d}
\end{figure}

\appendix
\section{Numerical convergence}
To test the issue of numerical convergence we have run a subset of the
previously described models with a higher grid resolution. 
In all simulations we use a torque cutoff of $r_{torq} = 0.8 R_{Hill}$
and (in the 3D cases) a smoothing length of identical extension.

The first set of simulations refer to the two-dimensonal setup (2D-disks).
From our standard resolution in 2D with $N_r \times N_\varphi = 128\times 384$ we first
doubled the number of gridpoints in each dimenension to a medium resolution of 
$256\times 768$ and then
increased it by again a factor of $\sqrt 2 = 1.41$ to $384\times 1152$, our high resolution case.
The results for an initial eccentricity of $e_0 = 0.3$, as displayed in Fig.~\ref{fig:aeres2d},
indicate first a qualitatively identical
behaviour for all cases, with only a minor shortening 
of the eccentricity damping timescale with higher resolution, which amounts to about 15\%
difference in $\tau_e$ between the standard and the HiRes case.
The differences in the migration can be attributed to the changing eccentricity damping,
as in the initial phase of the evolution the migration rate is nearly identical for all three 
models.
By chosing the highest value for the eccentricity we intended to pick out the most extreme
case, and we expect the variations to be smaller for lower eccentricities. 

Running a set of three-dimensional simulations on a typical parameter set with
$e_0 = 0.2, i_0 = 5^\circ$ with obtain the results displayed in Fig.~\ref{fig:aeires3d}.
Here, low resolution refers to $(N_r, N_\theta, N_\varphi) = (132, 40, 390)$, medium
resolution to $(N_r, N_\theta, N_\varphi) = (184, 56, 560)$ and high resolution to
$(N_r, N_\theta, N_\varphi) = (264, 80, 800)$, i.e. the three cases differ by about a factor
of $\sqrt{2}$ in the number of grid cells in each dimension, which gives about a factor of 16
in computation time between the highest and lowest resolution.
The results are very similar to the 2D case for the eccentricity and migration.
The eccentricity damping time scale shortens slightly (by about 10 \%)
when comparing the highest and lowest resolution runs, while the 
migration rate hardly changes at all.
The inclination damping time shortens as well upon increasing the resolution 
(Fig.~\ref{fig:aeires3d}), but this is largely a result of the change in eccentricity
damping time (see below).
Additional resolution studies performed in 3D for $e_0 = 0.2, i_0 =0$ and  $e_0 = 0, i_0 =5$
yielded very similar results and are not shown here. In the case with $e_0 = 0, i_0 =5$ degrees,
we found that the inclination damping time was hardly affected by the changes in resolution,
which is why the deviation in inclination damping times shown in
fig.~A.2 is actually due to changes in the eccentricity damping time.

Both 2D and 3D resolution studies indicate that there are still residual changes  at the
10 \% level
when the resolution
changes, and this change mainly occurs in the eccentricity damping rate.
Altogether our results demonstrate very clearly that the effect of an increased migration and
inclination damping for larger $e$ and $i$ in a locally isothermal disc
is a robust effect independent on resolution.
The absolute magnitude will depend on physical details in the vicinity of the planet, where 
physics that we have not included in our models (e.g. thermal and radiative effects, MHD
turbulence)
will play a role amongst others \citep{2004MNRAS.350..849N, 2005A&A...443.1067N, 
2006A&A...459L..17P}

\bibliographystyle{aa}
\bibliography{dirksen}
\end{document}